# Technical specification of a framework for the collection of clinical images and data

Alistair Mackenzie[1*], Mark Halling-Brown[1], Ruben van Engen[2], Carlijn Roozemond[2], Lucy Warren[1], Dominic Ward[1], Nadia Smith[1]

[1] Royal Surrey NHS Foundation Trust, Guildford, UK
[2] Dutch Expert Centre for Screening (LRCB), Nijmegen, The Netherlands

(*) Corresponding author: alistairmackenzie@nhs.net

## Abstract

In this report a framework for the collection of clinical images and data for use when training and validating artificial intelligence (AI) tools is described. The report contains not only information about the collection of the images and clinical data, but the ethics and information governance processes to consider ensuring the data is collected safely, and the infrastructure and agreements required to allow for the sharing of data with other groups.

A key characteristic of the main collection framework described here is that it can enable automated and ongoing collection of datasets to ensure that the data is up-to-date and representative of current practice. This is important in the context of training and validating AI tools as it is vital that datasets have a mix of older cases with long term follow-up such that the clinical outcome is as accurate as possible, and current data. Validations run on old data will provide findings and conclusions relative to the status of the imaging units when that data was generated. It is important that a validation dataset can assess the AI tools with data that it would see if deployed and active now.

Other types of collection frameworks, which do not follow a fully automated approach, are also described. Whilst the fully automated method is recommended for large scale, long-term image collection, there may be reasons to start data collection using semi-automated methods and indications of how to do that are provided.

## Glossary

**Anonymisation**: The process of removing or modifying personal identifiers from data so that individuals cannot be identified.

**Anonymised in Context**: Data is considered anonymised for a specific use case through the application of disparate secondary pseudonymisation techniques. By ensuring different datasets are pseudonymised using separate, unlinked methods, the risk of re-identification is reduced to a negligible level. As a result, third parties receiving the data cannot reasonably re-identify individuals, allowing the data to be treated as anonymised under relevant regulations.

**Artificial Intelligence (AI)**: The use of computer systems to simulate human intelligence processes, such as pattern recognition, decision-making, and predictive analysis. In imaging, AI may assist in image interpretation and risk assessment.

**Cases**: Individual instances of imaging events or diagnosis used in research, validation, or clinical assessment. Cases typically include imaging and associated clinical data.

**Client**: The person being imaged, including both patients and individuals undergoing health screening. Client is used throughout this report rather than patient, as it is more generic.

**Data**: Information collected during the imaging process, which may include clinical data, personal information, and imaging.

**Data Manager**: The individual responsible for overseeing the collection, storage, and management of imaging and associated clinical data.

**DICOM (Digital Imaging and Communications in Medicine)**: A technical standard for the digital storage and transmission of medical images and related information.

**DSA (Data Sharing Agreement)**: A document to set out the purpose of the data sharing, cover what happens to the data at each stage, set standards and help all the parties involved in sharing to be clear about their roles and responsibilities.

**Ethics**: A set of moral principles and professional guidelines that govern actions for the purpose of this document specifically in imaging, AI validation, and data handling to ensure client rights and welfare are protected.

**Encryption**: The process of converting data into a coded format that can only be accessed by authorized users with the correct decryption key.

**EPR (Electronic Patient Records)**: A system used by healthcare providers (e.g. in the UK by the NHS) that brings a patient’s medical records together in a single, secure digital location. It allows healthcare staff to access everything they need to know about a patient, improving patient care and staff efficiency.

**Events**: Medical system encounters, such as imaging appointments, medical examinations, or interventions, recorded as part of a client’s screening or diagnostic journey.

**Ground Truth**: The definitive medical outcome of a case, typically established through biopsy, long-term follow-up, or expert consensus. This may differ from the initial diagnosis or AI predictions.

**Images**: Digital radiographic representations, such as mammograms, used for breast screening, diagnosis, and AI validation.

**Image processing**: Methods used to perform operations on an image to enhance it or to extract useful information from it.

**Information Governance**: The framework of policies and procedures that ensure sensitive client data is securely accessed, stored, shared, and protected from unauthorised use.

**JSON (JavaScript Object Notational)**: An open standard file format and data interchange format that uses human-readable text to store and transmit data objects consisting of name–value pairs and arrays (or other serializable values).

**Licensee**: An individual or organization granted permission to use shared imaging data under specific ethical and legal agreements.

**NBSS (National Breast Screening System): The** IT system set up by the NHSBSP to manage the screening programme. It is used to book appointments, and track screening histories.

**ODBC (Open Database Connection)**: is a standard Application Programming Interface used for accessing database management systems.

**Opt-out**: The process by which a client chooses to withdraw consent for their data to be collected, stored, or used in research or validation studies.

**OMI-DB (OPTIMAM Mammography Image Database)**: A database that collects NHS Breast Screening Programme (NHSBSP) images from multiple breast screening centres and breast clinics across the UK and has been created to serve as a large repository of de-identified medical images to support research involving medical imaging.

**PACS (Picture Archiving and Communication System)**: A medical imaging technology used to securely store, retrieve, and share digital radiology images.

**Patient:** A person undergoing medical treatment or has a health issue under investigation.

**Personal Data**: Any information that can be used to identify an individual, such as name, date of birth, health ID number, or identifiable imaging data.

**Pseudonymisation**: A data protection process where personal identifiers are replaced with unique codes to reduce the risk of identification while maintaining data integrity for analysis and validation.

**RIS (Radiology Information System)**: Software that stores and manages medical image data.

**Site**: A physical location, such as a hospital, research center, medical clinic, or specialized healthcare facility, where clinical activities are carried out. These activities include enrolling participants, administering investigational treatments or interventions, collecting data, performing patient monitoring, and adhering to research protocols.

**Secrets**: Any sensitive data an application needs, including credentials, tokens, and private keys.

**SMART Box (Secure Medical Anonymiser for Research Trials)**: A secure system used for anonymising and managing medical images for research purposes while maintaining regulatory compliance.

**Validation**: The process of assessing the accuracy, reliability, and clinical applicability of AI tools by comparing their outputs against ground truth data.

# Contents

# 1 Introduction

The aim of this report is to describe the characteristics of a collection framework for radiological images for use when training and validating AI tools. This includes not only the collection of the images and clinical data, but the ethics and information governance processes to consider to ensure the data is collected safely, and the infrastructure and agreements required to allow for the sharing of data with other groups. There are many good reasons to set up research databases of clinical images, but it must be borne in mind that considerable work is required to set up the collection process. There are a number of publications setting out the methods and challenges of collecting images for research purposes [1], as well as describing the challenges of extracting data from a Picture Archiving and Communication System (PACS) [2].

The primary purpose of this report is to describe methods and processes for collecting images and clinical data for tuning, training, testing and/or evaluating AI products, but there may be other purposes for collection of the images. This document is intended to provide a comprehensive outline of the purpose, operation, methods, policies and governance of collecting and using images. The focus of the document is on how the images and data will be collected.

Before the collection process is started, there are some decisions to be made as they influence the setup of the data collection process, which might be difficult to change later. The choices will depend on the purpose of collection and availability of data sets. Some examples of choices are as follows:

- Modalities from which the images will be collected: collection of planar images (mammography, general radiography, dental), 3 dimensional images (computed tomography (CT), magnetic resonance imaging (MRI), tomosynthesis), multi frame images (ultrasound, fluoroscopy).
- Collection of unprocessed images alongside the processed/reconstructed images.
- Collection of additional client data from other sources (e.g. cancer registry, or data from e.g. genetic tests performed and stored separately).
- Numbers: collection of images/data from all clients, random selection or choice of sub-set.
- Direct acquisition from the imaging systems or PACS: generally, it is easier to collect from PACS rather than each imaging system. If unprocessed images are not stored on PACS, then acquisition from the imaging system may be required. How and where to link images/data from different sources.
- Client consent: assumed or explicit consent.
- Timescale: set period or continuous collection.
- Setup the anonymisation process.
- Single or multi-institution collection.
- Data integrity checks.
- Data cleaning process.
- Share the data with other institutions or not.
- Acquisition of follow-up information.
- Annotate images to demonstrate regions of interest and type.

Particular attention is given in this document to how the data is de-identified at the point of collection by automated processes. Often it is necessary to be able to update a client's clinical data in the database in the future if they have returned for further clinical examinations. Therefore, it is not possible to simply anonymise all fields, as it will not be possible to identify which records to update. To overcome this, a double pseudonymisation process can be used. Within the clinical site the data is pseudonymised i.e. client identifiers are replaced with a pseudonym. An extra layer of security would be the images and data are pseudonymised a second time in a second location with different controllership relations, such that the data within the image database is rendered anonymised in context to third party researchers, as they have no access to the original data or either pseudonym lookup. Further challenges are the collection of clinical data associated with the imaging and also to link to other imaging events that may have been undertaken on the same client.

The dataset collected should be representative of the different populations and equipment that an AI tool encounters clinically. Additionally, there should be large enough numbers in smaller subgroups of the populations. Therefore, it is vital that any data collected for training or evaluating AI is from multiple sites and is large scale.

It is important that when data is shared with third party groups, that the purpose for which it is being used is of clinical benefit as outlined within the ethical approval, and the data and images are tracked so it is clear who is using which data. This will require processes for the sharing of images. This document describes such processes in Section 7.

# 2 Protocol

## 2.1 Collection Processes

The first principle of image and data collection systems is that all data that is transferred from the clinical sites to the central image database is fully de-identified (Section 2.3) and that at no point do researchers have access to client identifiable information.

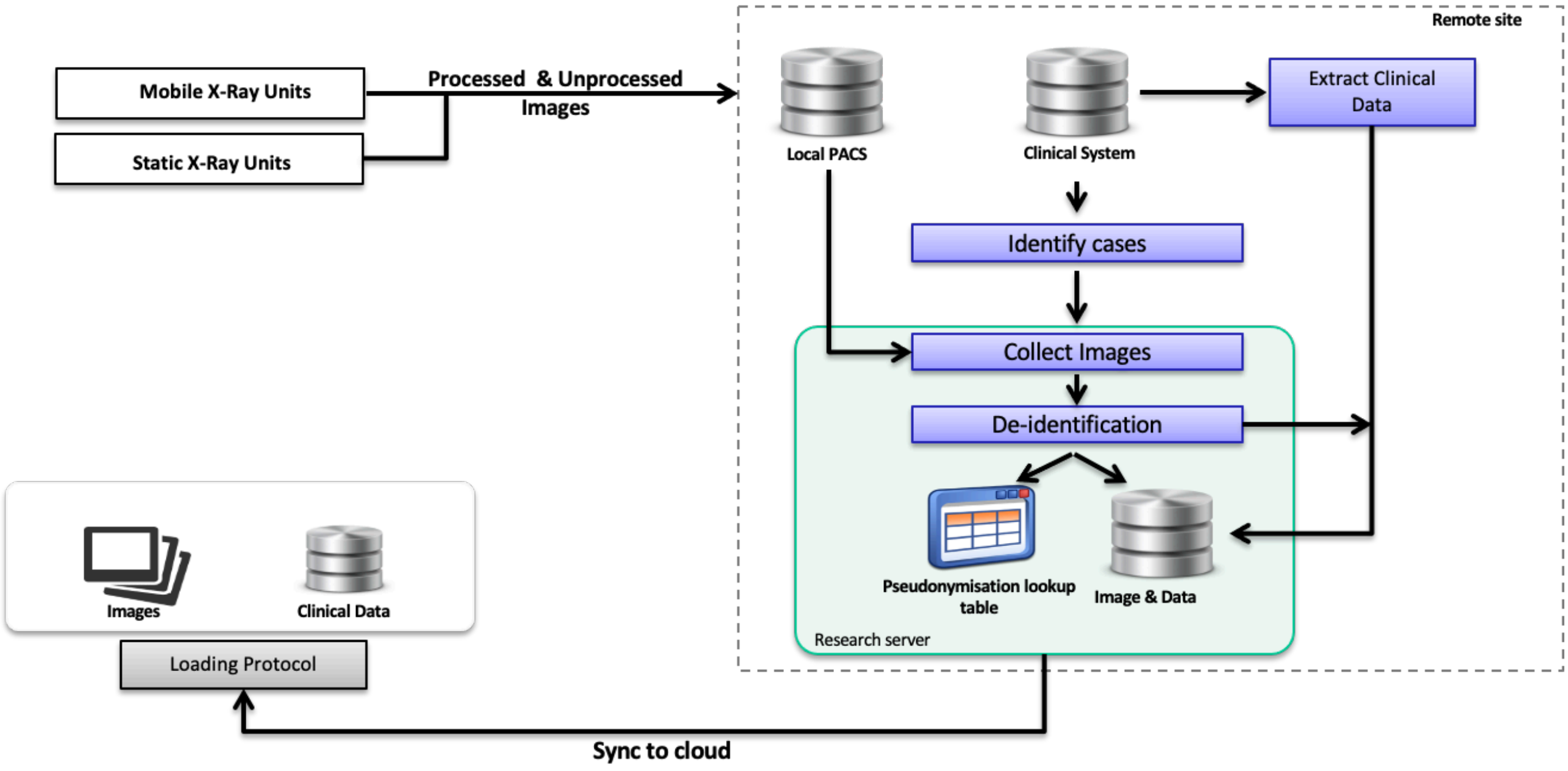

**Figure 1: Example of the image collection process for mammography images in the UK Breast Screening Programme** *[1]*.

### 2.1.1 Automated data collection

The following section describes an automated collection system that can be deployed on a large-volume collection site. Each site can have a dedicated research server on which the automated system is deployed. This dedicated server is typically a virtual server provisioned within the imaging department's infrastructure and managed by their information technology (IT) department. More information on the research server operation is provided in Appendix 1, in which an example of such an automated collection system – the Secure Medical Anonymiser for Research Trials (SMART) box developed by the Royal Surrey NHS Foundation Trust in the UK – is given.

All non-anonymised data must be dealt within the server and remains within the clinical site. Access to the research server is restricted to approved data managers with permission of the site to access the data. The permission should be in writing and set out what can be accessed. The amount of contact that the data managers have with any client data must be kept to a minimum. This is achieved by implementing an automated collection system. This allows the collection to continue without any contact with client information by the data managers. The exception to this is the minimal contact required to set up the pseudonymisation process and to test and verify that it is working correctly.

The collected images are typically in the Digital Imaging and Communications in Medicine (DICOM) format. However, there are modalities that store the images in jpeg format. If these images are required, then the system must be set up to collect both types of images.

The associated clinical data needs to be collected in addition to the images. The data will be stored on a clinical system, such as Radiology Information System (RIS) and Electronic Patient Records (EPR). Assuming that a suitable clinical system is available and able to be queried then the SMART box collection process can be driven by data extracted from this clinical system. Connection to the clinical system is made between the SMART box and the site's clinical database. This is achieved through a direct connection to a database or system backend (e.g. through an open database connection (ODBC)), through standardised healthcare communication protocols or via some other connection protocols. It is vital to focus on connections and data collection processes that can be automated, scaled and ongoing.

The automated collection system uses the clinical system connection to regularly query the clinical database. Criteria can be set here on what cases to collect e.g. positive cases or a selection of normal cases. There may be further criteria that can be applied to ensure key sub-groups are sufficiently represented. Once a case has been identified, the next step is to extract the images from the site's PACS or imaging system to the onsite research server. A query/retrieve connection needs to be implemented; it may need assistance from the PACS vendor to be able to implement it. This connection is established between the site's PACS and the research server. In effect, this allows automated extraction of images from the site's PACS to the data collection server, based on input from the identification stage. At this point, the images are not yet de-identified. The automation of this step has many benefits, including the ability to run the collection process overnight to avoid any burden on the PACS or local network.

The images for the cases identified are pulled from the site's PACS using an automated query/retrieve command. These not yet anonymised images are received by the research server Digital Imaging and Communications in Medicine (DICOM) receiver. This process is automated and does not need any human intervention.

The images received by the research server are then pseudonymised, and the pseudonym and the key for the pseudonymisation is saved in an encrypted form in a local database residing on the research server. This database stores the encrypted client ID number, the internal episode ID and Study unique identifier (UID) of the cases. Details of the DICOM tags that are de-identified and those that are pseudonymised are given in Appendix 3.

The clinical database connection is used by the research server to extract information about the case. The information is pseudonymised using the lookup table already populated (see Figure 1) - it is recommended to encode the data in a JavaScript Object Notational format (JSON) format or some other machine-readable format to allow consistency in data encoding between sites and to facilitate ontological descriptions and software libraries to be created to parse the data. These JSON formatted files should be placed in the same folder structure as the images.

### 2.1.2 Semi-automated data collection

The fully automated method is recommended if large scale, long-term image collection is to be undertaken. However, there may be reasons to start data collection using semi-automated methods.

In this case, it is possible to connect a collection device, such as a hard-drive or server, to the imaging system. These will require time to be spent switching devices and copying files to the main server. It is important that the images and data are anonymised in the same way as previously described. There may be added security issues with devices, as they may be stolen or lost and so they must be securely encrypted. There can be problems in copying data to hard-drives and checks may be required to ensure all the imaging data has been copied.

#### 2.1.3 Unprocessed images

The need to collect unprocessed/projection images is important for many research applications. Unprocessed images are essential for studies on different types of image processing/reconstruction and computer aided detection. The pixel values and the noise in the unprocessed/projection images also have a direct link back to the physics of the image formation. Once the images are processed this link is lost and cannot be recovered. Unprocessed images can also be used to study the effect of different imaging systems' design parameters on clinical performance.

Unfortunately, most PACS systems store only processed images, and the unprocessed images are discarded. The unprocessed images are generally discarded due to storage costs rather than any technical reason.

The most efficient method to collect unprocessed images is for these images to be also stored on PACS. However, it may be possible to collect images directly from the imaging systems. It will be possible to collect unprocessed images retrospectively from a PACS, but images on an imaging system will be deleted after a certain amount of time. The processed images should always be available from PACS.

If non-DICOM images are also to be stored e.g. jpeg with ultrasound, then there needs to be a way to store technical details about the system and settings in a similar way that DICOM data is stored.

### 2.2 Addition of ground truth and additional data

For an image database to be useful, it must contain outcomes from the imaging and ideally clinical outcomes from further tests such as biopsy, as these will be considered the ground truth. The ground truth will mostly come from querying the clinical systems. The image interpretation and any further clinical tests will be available days to weeks after the images are acquired. The collection system will therefore need to periodically query the clinical system for updates. A system will be required to be able to link the clinical data to the anonymised case, such as a look up table linking the clinical IDs with the anonymised IDs. The collection system will then send this data to the main database and update its records. It may be possible to connect to other databases containing demographic data such as race or socio-economic factors.

The querying for ground truth should be taken over a longer timescale than weeks as there may be further clinical events such as imaging or other clinical exams. The system needs to be able to link further acquired images and update the clinical information. Initial diagnoses that are subsequently found to be incorrect are especially useful for the database and need to be rectified otherwise any AI developed or tested using this data will further enhance the error.

For most purposes, the clinical system will provide sufficient information although, depending on the studies required, some of the relevant information may not be available. This may require experienced readers to transfer some information (e.g. from clinical notes) to the database and ideally to accurately mark the size and location of areas of interests e.g. a lesion. An example of marking further information is shown in Figure 2.

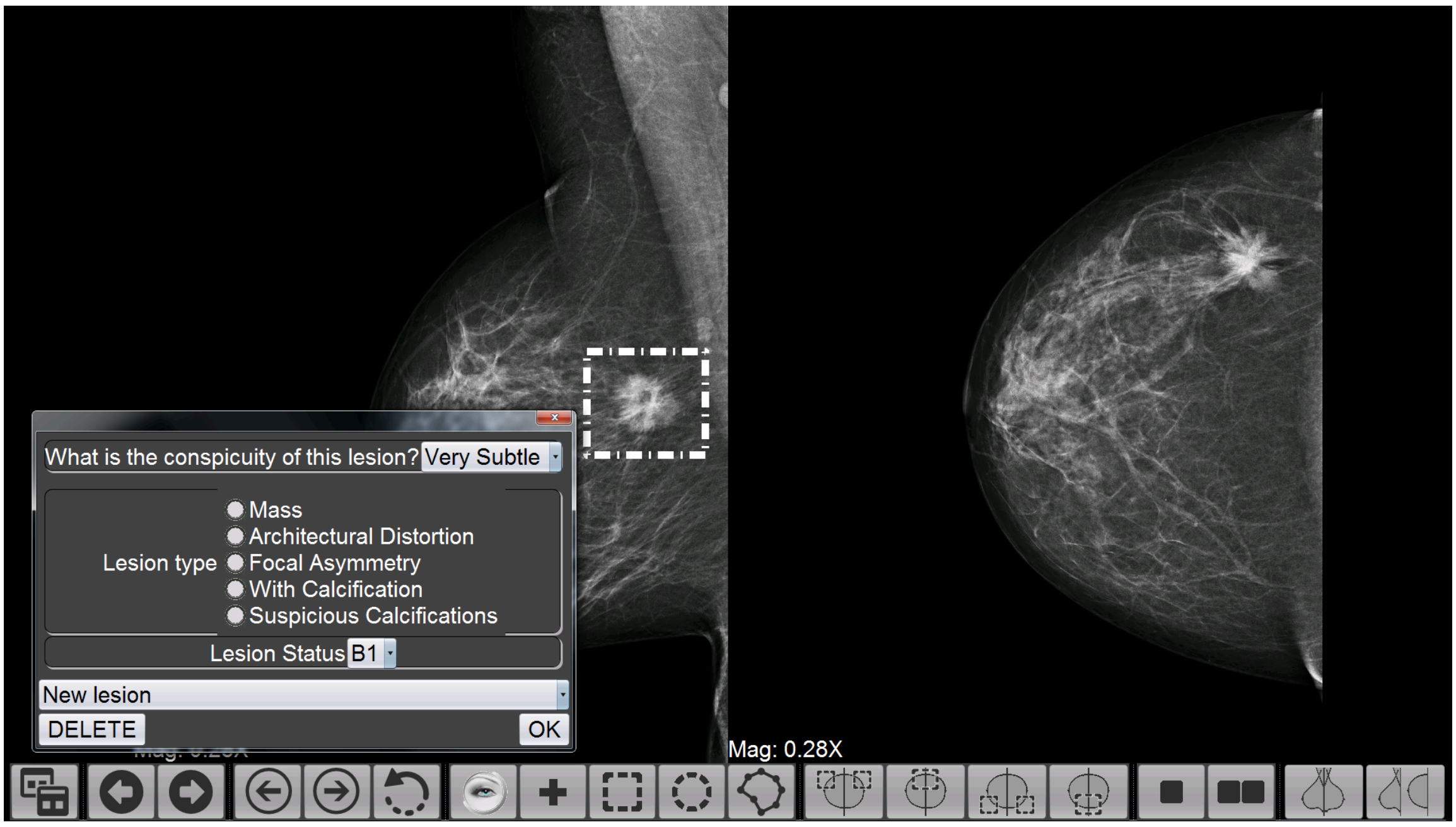

**Figure 2: Example of marking of cancers in a mammography image.**

## 2.3 De-identification Processes

Each medical image is stored in a standard DICOM format. In addition to the actual image, there is a DICOM header with information stored in fields that are known as “tags” which contain a very large amount of information including client identifiable information. In order to preserve the confidentiality of the clients the following changes can be made to each DICOM header at the clinical site before it is transferred to the central server.

The de-identification must occur before they are used or seen by any data administrators, or researchers outside of the imaging department. The data may be deleted or replaced by another value (pseudonymised). The following are some examples of pseudonymised tags:

- 0010,0010: PatientName: This is replaced with a code, the use of an auto incrementing number is the simplest implementation.
  N/B: The auto incrementing can be global (i.e. across all institutes) or local (each institute has its own auto-incrementing number)
- 0010,0020: PatientID: This must be changed, it is simplest to make this the same as PatientName.
- 0010,0030: PatientBirthDate: This must be changed, however, it is useful to know the patient's age, so often the year remains the same, but the day and month are set to “01”.

A full set of tags that could be de-identified are listed in Appendix 3. DICOM Supplement 142 [3] could be a useful reference for deciding on tags that need to be de-identified. There are tags that manufacturers allocate, and for which content is not defined in the DICOM standard, and therefore there is a risk that private tags will contain personal data and the DICOM supplement recommends de-identifying these tags. However, the tags may also contain useful information for the researchers and so a collection process may decide to not de-identify certain tags, but careful checking must be undertaken as a private tag may contain useful information for one imaging system but personal data in another.

The acceptability of the pseudonymisation procedures will need to be checked that the de-identified data is without personal information. It may be useful to take advice from an Information Governance Officer on the acceptability of the procedures.

There may be personal information (both client and radiographer) burnt into the image as well. A collection process will need a method to detect them and erase these details from the image or exclude the images from collection.

### 2.3.1 Data Cleaning and Staging

Time needs to be spent on checking the accuracy of the data. It may feel automatically acquired data will be accurate. However, there are a number of issues that may arise. Some of the data may be incorrect at source, the linking of images to clinical data may be incorrectly made or missed, the same client may appear twice in the data, incomplete data may be received, imaging from other modalities may be received accidently or there may be multiple names given to the same imaging procedure.

To fully curate a dataset will be time consuming and the level of curation required should be considered. Normally a curation pipeline would be implemented to undertake the following steps:

1) Transfer the collected images and data from the staging area to the 1st stage storage
2) Undertake the secondary pseudonymisation steps to replace the original pseudonym with a secondary pseudonym, rendering the data anonymised in context (see below)
3) Scan pixel data to ensure technicians initials or other identifiers are not present and black them out if they are
4) Identify and flag any digitised scans or other erroneous data (e.g. dose reports disguised as images or measurements values stored in ultrasound images)
5) Produce an image linkage file to capture the Study, Series, Image relationships within a client's data
6) Create a new version of the clinical data (in JSON format) with only the properties deemed suitable for use by researchers included
7) Push the resultant version of images and data to final cloud storage

### 2.3.2 Anonymised in Context

Anonymised in context is a concept where data is considered anonymised for a specific use case. This is commonly achieved by implementing further levels of anonymisation (secondary or even tertiary levels) and is called secondary pseudonymisation techniques. The methods of anonymization must be unlinked. This results in a set of images and data that cannot be directly related back to the originally anonymised images and thus even more remote from the originally acquired data.

These processes make re-identification so unlikely that, from a third party's perspective, the data is effectively anonymised. Even if one dataset is compromised, then it does not reveal the full identity of individuals due to the segregation of identifiers. The data controller might still have a theoretical ability to re-identify (via key retention), but third parties receiving the data do not have access to linking mechanisms. These processes align with UK GDPR and Information Commissioner's Office (ICO) guidance on anonymisation, where data is contextually anonymised if re-identification risk is negligible for the recipient.

These processes are particularly useful if the data is to be shared externally or other internal projects.

## 2.4 Storage and backup

The images and data need to be stored securely, this could be a research server, but for scalability and security and for large volumes it may be better to save them in the cloud. The laws of each country may limit the location/region of the object storage cloud solution that the data are stored in.

Considering the effort to collect images and data, there must be a backup of the data, this may be easier on the cloud.

## 2.5 Client and Public

The image and data collection may be collected using explicit or assumed client consent. Information governance of an organisation or national laws and regulations may prohibit the acquiring of data without client consent. If it is possible, then a decision to acquire client consent or not is required. Generally, for the number of images required for evaluating AI then client consent may be prohibitively expensive and time consuming. There are three main criteria required to acquire images without consent: no personal data to be shared, information available for clients, and an option to opt-out.

### 2.5.1 Explicit client consent

To acquire client consent for the collection of images and clinical data, there must be information available to the clients describing the process of image collection and the purpose of the use of the images and data, including what can be shared. This can be sent out in leaflet form with any invite for the examination. Clients will also receive a form for them to agree or not give permission for the images to be collected. If the form is not completed or the client requires more information, then time may need to be spent with them to explain the process and organising the signing of the form, often this will be done by the radiographer or a person who can be contactable (by phone or email).

The records of consent must be kept and implemented and ensure that the request of the client is upheld. There should be a process for a client who changes their mind (within a reasonable time scale).

### 2.5.2 Assumed client consent with option to opt-out

If images and data are acquired without client consent, then it is necessary to inform clients that their data may be collected. For example, the display of a poster or the availability of leaflets may be sufficient for information, however, some countries may have stricter requirements e.g., informed directly during the imaging invite. Any posters should be easily seen and should be easily understood by members of the public. The posters could be in the reception, waiting area and/or imaging room. The clients need to be informed that image collection is occurring, and that no personal or identifiable data will be used, the purpose of image collection and how to opt-out if required. There should be further information available for the clients, this could be in the form of a list of frequently asked questions (FAQ). Normally it is vital to ensure that the collection sites' privacy policy and statements are suitable and include the use of de-identified data for research.

There will need to be a group of people who run and control the database and ensure its security and smooth running. Some of the staff involved are listed in staff roles (Section 2.6) below, it is also vital that a representative from a patient group is included, to give a client's perspective on the work.

If client consent is not undertaken, it may be necessary to have a process for the client to opt-out of the data collection, i.e. clients can opt-out of their data being processed into anonymised forms for collection into the image database. If there is a system of assumed consent, then there are two methods of opt-out:

- National opt-out: If there is a national or regional opt-out scheme where clients can register their request to opt-out.
- Local opt-out: requires the client to approach a member of staff to request they are opted out. This member of staff would then need to contact the assigned local database contacts (usually a PACS team member) to register that client's hospital/health number on the local database.

During every collection run, the system must consult this opt-out list and compare it with the new clients being proposed to be added to the database and exclude any that match. If a client has selected to opt-out, an automated process prevents data being collected and also checks if this client's data has already been collected (through matching hashes). If so, a deletion cascade is initiated by which the client's information is removed for the local pseudonym lookup and the image database.

### 2.6 Staff Roles

There are several roles that need to be considered as part of ensuring that data and images are collected securely. These are examples of some of the roles, but they may vary between countries and local organisation structures:

- Database manager(s) - these are the technical staff responsible for setting up and maintaining the collection systems.
- Site lead – Each collection site should have a clinical lead that is involved and advocating for the research database and collection.
- Senior Information Risk Owner- Staff member at the institute, which claims controllership or ownership of the research database. Their role is to sign-off the processes, policies, and agreements.
- Information governance leads – they will exist at the central organisation owning the database and at each of the collection sites. They will advise on the acceptability of the proposed processes and agreements.
- Clinical information governance lead - A clinical member of staff will enforce the region's general overarching data handling rules from either a clinical point of view or standpoint (e.g. Caldicott in the UK).
- Main contact within the imaging department for clients to approach for further information.

### 2.7 Software for image collection

There are layers of different software programs, which effectively connect into pipelines to run the collection processes. There are a number of these programs available for free or off the shelf that can assist, but there are not simple pieces of software that can accomplish the full image collection, preparation and pipeline activities. For example, for the OPTIMAM Mammography Image Database (OMI-DB) [1, 4], all the collection functions have been bundled into a SMART box and system (a full description is in Appendix 1). The software functions for establishing such a database and some options for software, which provide this function are listed as follows:

1) Clinical system integration is a process to integrate with and extract data from different healthcare information systems. This could take many forms, so is likely to be a bespoke script/software; however, Trust Integration Engines (TIEs) [5] could be a starting point.
2) DICOM Query/Retrieve (Q/R) – The ability to query a PACS and route the resultant images to a receiver. There are many existing libraries for achieving this (e.g. DCM4CHEE [6]/DCMTK [7]).
3) DICOM Receiver – The ability to listen for and receive a DICOM image of DICOM networking protocols. There are many off the shelf products, but ideally a command line interface (CLI) is required (e.g. DCM4CHEE [6]).
4) De-identification – The ability to de-identify the incoming images and clinical data to a specific specification and write the pseudonym lookups to an encrypted database. This is likely to be a bespoke script/software.
5) Transfer to cloud or local network storage – The ability to move the data to local storage or the cloud utilising collect authentication and encryption methods. Cloud application programming interfaces tools like Rclone [8] can assist.

6) Data pipelining and curation in the cloud – The ability to move data from staging areas to production, curate and prepare and move to storage is likely to be bespoke and ideally is developed with cloud integration in mind.
7) Querying data for dataset selection – The ability to query the data in order to be able to characterise and identify a specific dataset. This will likely be bespoke or some form of statistical package.
8) Sharing data with third parties – There are various methods by which this can be achieved which will vary based on the ethical approval. In general, a bespoke system with secure secrets sharing and audit logs will be required.

### 2.8 Costs and funding for image collection

The overall costs of collecting the data will depend on the image collection process and the number of sites and images to be collected. Costs should not be under-estimated. They include not only staff time and equipment, but there may be costs within the imaging unit, especially if they store unprocessed images on their PACS as part of the processes. There will be image storage costs and if the cloud is used then there can be image transfer costs as well. Cloud storage costs are not negligible and will continue for the whole length of any project.

Costs may include:

- Staff time: computing, PACS, radiology staff for setting up, running, curation of data, general administration.
- Equipment costs: hardware.
- Equipment maintenance
- Software: either purchasing software or developing software.
- PACS: extra costs if unprocessed images are stored.
- Storage costs: either on a server or the cloud.
- Transfer costs: it may cost to move images and data around.

It may be possible to receive funding as part of a research grant or funding from industry, though these are likely to be temporary and not necessarily cover long-term storage costs. Long term funding may require a model of licensing the use of the images collected to researchers and industry. In these cases, then the licensing fees can be used to cover costs from imaging departments.

## 3 Ethics

The ethics of the processes listed here need to be considered and actioned. This section covers the collection and the sharing of images and data. The two are closely linked as data cannot be gathered without having processes that ensure the data is shared and used within a strong ethical framework.

The ethical requirements of any project must ensure that any potential consequence of a project ensures that the risk of harm such as personal data entering the public domain is kept to a minimum. The processes that are set up must be carefully considered to ensure that ethical principles such as privacy and fairness (i.e. the data is collected ethically and without introducing bias) are followed. The collection of image data will require some form of ethical approval at a national or regional level. The exact methods for obtaining ethical approval will be country dependent, there may be variations in information required for an application and procedures that need to be put in place to meet the ethics requirements. Engagement with the ethics process should be undertaken as early as possible in the setting up of a collection process.

The main considerations from an ethical standpoint in data collection will be around informing the patients, ensuring good data security and that the data is used ethically. Most of the information required and decisions to be made are discussed in this report.

The ethics approval body will likely require some reporting of progress to monitor the safe data collection and appropriate use of the data.

Ethics approval is normally sought from a central body responsible for ensuring and enforcing the ethical use of data. In general, these bodies will need clear evidence of how the project is conforming with the five safes:

**Safe data:** that data is treated in such a manner to protect any confidentiality concerns.

**Safe projects:** research projects are approved by data owners for the public good.

**Safe people:** researchers are trained and authorised to use data safely.

**Safe settings:** a safe and suitable environment that prevents unauthorised use.

**Safe outputs:** screened and approved outputs that are non-disclosive.

For example, in the UK, the Health Research Authority (HRA) is the responsible body for ethical approvals for research studies. HRA has a concept of a 'research database', which is a special form of ethical approval given to initiatives looking to form databases for the purpose of sharing data to assist and stimulate research. One of the most important criteria is that a robust and auditable data sharing process, including a data access committee, is set up and in place as long as the research database exists. The data access committee will need to ensure that the projects that data is shared with follow the Safe principles and have their own ethical approval.

# 4 Site Engagement

## 4.1 Setup

### 4.1.1 Access to clinical images

In order to permit the data managers to set up and manage the collection process, then permission to access the data will need to be obtained at each collection site. This permission must be in writing and set out what can be accessed and can sometimes be called a 'letter of access'. The requesting body must ensure that data managers are aware of their responsibilities when handling personal data and that any information governance training has been completed. Typically, these permissions should be time limited.

### 4.1.2 IT/PACS managers

The implementation of the collection process described in Section 2.1 may require considerable work by the data collectors, IT/PACS managers, and lead clinical staff. They need to ensure that they have sufficient power in the hardware. The hardware and software must fulfil the following steps:

- Client and data management through an internal web portal.
- DICOM image collection and de-identification through a DICOM server.
- Data transfer to the image database - they will also need to ensure that the transfers are secure.

### 4.1.3 Informing clinical staff

The clinical staff need to be informed of the image collection process; they will often be the main person that the patient will ask any questions about the collection and/or opt-out. They will need instructions on how to answer these kinds of questions and a leaflet explaining the image collection, which can be given to the client/patient asking for more information and a list of frequently asked questions. There will need to be a contact for questions that go beyond the radiographer's knowledge.

## 4.2 Governance

### 4.2.1 Data protection regulations

All countries must have regulations to cover the safe use of data and ensure the privacy of its citizens. The General Data Protection Regulation (GDPR) [9] is a European law (also directly applicable to UK law) that covers the following relating to the processing of personal data:

- Lawfulness, fairness and transparency.
- Purpose limitation.
- Data minimisation.
- Accuracy.
- Storage limitation.
- Integrity and confidentiality.

### 4.2.2 Data Protection Impact Assessment

Under the GDPR, a Data Protection Impact Assessment (DPIA) must be completed whenever there is a change in an existing process or service, or new processes of an information asset is introduced that is likely to involve a new use or significantly changes the way in which personal data is handled. The DPIA is a process that assists in identifying and minimising the privacy risks. A useful source of information and a template can be found on the EU website [10].

An effective DPIA is a tool (a Risk Assessment), to help identify the most effective way to comply with Data Protection obligations and meet individuals' expectations of privacy allowing the organisation to identify and resolve any problems at an early stage, reducing the associated costs and damage to reputation which might otherwise occur.

With regards to image collection terms may be defined as:

- Informational privacy – the ability of a person to manage their information and its use.
- Confidentiality – is the right of an individual to have personal, identifiable medical information kept private. Such information should be available only to those with a legitimate right of access.
- Information Security – relates to protecting information and information systems from unauthorised access, use disclosure, disruption, modification, perusal, inspection, recording, destruction.

## 4.3 Contracting/Agreements

The collection process needs to be fully documented. This allows proper audit of the processes. An important part of this is the contracts and agreements between imaging centres, image collectors, data controllers and those who are using the data. These will govern the rights and obligations of all the parties involved.

The main agreement is the collaborative agreement between the parties in the image collection process. The agreement can cover the following:

- intellectual property rights and any commercialisation of products developed using the image collected,
- clarity on ownership of the data,
- duration and termination processes for the contract,
- liabilities and indemnities,
- how to add new parties,
- any financial arrangements,
- publications and announcements,

- confidentiality, for example proprietary information,
- dispute resolution,
- data protection.

It may be necessary to add new members to a collaborative agreement; in these cases it may be more practical to use an adherence agreement rather than have all existing partners re-sign. An adherence agreement would set out that they agree to follow the original agreement.

A Data Sharing Agreement (DSA) will be required which may be combined with the collaborative agreement, but in most cases, separate documents may be easier. The DSA sets out the agreement for the imaging department to allow their data to be collected and shared and sets out the responsibilities of the centre creating the image database. The DSA needs to include the data protection agreement between the parties. This includes the terms of the agreement and purpose, set out types of data that can be shared and the anonymisation, obligations during processing of personal data. It will set out the data to be shared e.g. de-identified images, clinical data, racial and ethnic details. The processes to be undertaken at the end of the contract need to be set out. The DPIA will need to be undertaken before the DSA is agreed.

If required, then further agreements are needed to share the image and data to third parties. The data access licences, the data controllers, and the researchers are described in Section 7.

All of the agreements need to be legally robust, there are websites that produce template agreements (e.g. [11]), but the responsibility for ensuring that they are legally sound will lie within the organisations. This may be more challenging when sharing images internationally. A legal team can help with the writing of agreements. Within these agreements, there must be consideration to timescales for the arrangements and any conditions for terminating the arrangements.

### 4.4 IT/PACS checklist

Before setting up a site, the local IT and PACS teams must discuss the placement of a data collection server and its interconnectivity requirements. Good co-operation between the local and collection parties is vital to the success of data collection. This should be undertaken as early as possible. It is good practice to have a site setup guide and checklist prepared for these interactions, examples of these can be seen in Appendix 1. IT will be responsible for provisioning the server, allowing access, cybersecurity reviews and opening the required exceptions in the relevant firewalls. The PACS team will be responsible for onboarding of the new server as a DICOM Q/R node on the site's PACS.

It is common to expect that IT will have several forms and change processes to comply with, such as e.g. New System process within the UK's NHS, 3rd party contractor onboarding processes and cyber reviews. Ideally, the IT team will take ownership of the research server and be responsible for ongoing cyber and updates.

If external drives are used in the process, then IT will need to clear the security of these devices.

## 5 Data collection site setup

The onboarding of the clinical sites is the most important aspect of the image collection process. In order to ensure a large-scale and representative dataset, it is vital to be able to collect from a large number of sites, hence the onboarding processes need to be well defined and robust. Once the initial engagement is complete and agreements are completed, then the actual process of setting up the systems can proceed.

### 5.1 Provision of a software collection server

Typically, a virtual machine will be provided by the hospital's IT department according to the specification in the site setup guide (see Appendix A1.1) provided to them. When access to this server is made available, the central team can remotely log on and deploy and configure the collection systems.

### 5.2 Collection System Setup

Once access to the host server is provided, the collection system can be installed and configured for that site. To ensure scalability it is vital to implement processes that allow configuration of sites through automated integration deployment methods which allow placeholders to be replaced by configured options automatically upon deployments (this allows automated integration and updates of configurations). This allows the deployment of secrets - that is, the process of securely distributing and managing sensitive information - in a secure manner and the alteration of configuration to be managed centrally.

During the engagement phase, details of the various configured properties should have been gathered, and these are entered into the checklist documents (see Appendix 1).

### 5.3 User Acceptance Testing

Testing is vital to ensure that the systems are operating correctly. In addition, User Acceptance Testing (UAT) should be undertaken with the local sites team to ensure the following works satisfactorily: systems integration, the results of the de-identification, and the burden on the network traffic or source systems due to the data collection. Data collection should not slow down or interfere with the local clinical practice. An example UAT can be found in Appendix 1.

### 5.4 Maintenance Plans

To ensure that the responsibilities for maintenance of the research server are clear and assigned, a maintenance plan is desirable. This document defines the maintenance processes for a deployed automated data collection (e.g. in the form of a SMART box) including the frequency of updates, monitoring plan and update process. An example maintenance plan can be found in Appendix 1.

The maintenance plan should:

- Include contacts.
- Ensure software utilised by the server is up to date.
- Ensure that all connections are still working.
- Include a proactive and corrective process involved in the maintenance of the image collection.

The central team can adopt a medium risk appetite in relation to updates to the automated data collection by ensuring that urgent/security patching are undertaken urgently, however, routine version updates which do not contain urgent security patches will only be rolled out when new features are required. This medium risk approach balances the need to ensure that the SMART box is running on a stable platform (known working version) with the need to patch urgent security flaws.

## 6 Advisory Board

It is recommended to have an Advisory Board (or Steering Committee). These will be set up for guidance and direction of the image collection, sharing and research undertaken. Typically, this group should contain not only clinical staff, but also data scientists, computing scientists, medical physicists and patient representatives. An example of a Steering Committee is referenced in Appendix 2.1.

# 7 Processes for third party access to data

Sharing of the data to third parties is the main reason for data collection, hence the sharing processes need to be streamlined and as automated as possible. Depending upon the regional rules and digital maturity, a Trusted Research Environment (TRE) [12] or Secure Data Environment (SDE) [13] may be mandated. In these cases, the data is stored centrally for researchers to work on where they can access the data but only extract their results but not the original data. Depending upon the maturity of the region's TRE/SDE and their ability to host large imaging datasets and provide GPUs, use of centrally accessed resources may not be feasible. In these instances, data can be shared with 3rd parties to host on their own environments given certain assurances on the infrastructure security and access controls.

## 7.1 Data Access Committee

The collection group will need to set up a Data Access Committee (DAC) to decide on who and how other groups can access the images. This access group will need to have a mix of roles and expertise. Some example documentation for a DAC can be found in Appendix 2. In general, the DAC sharing process should be:

1) A third party applies for access using an application template.
2) An administrator makes a first pass of the application to ensure specific criteria are met (such as ensuring CV is present, Principal Investigator (PI) is suitable, details on ethics and auditable storage are suitable). Criteria will need to be set out in advance.
3) The application is circulated to the committee and individuals complete an access committee feedback form.
4) Feedback forms are collated by administrator and, when quorate, are reviewed. If there are a number of "Require More Information" tagged on the forms, then the administrator provides feedback to the 3rd party to request a resubmission.
5) A final decision is taken to permit access or not.

## 7.2 Data access licence agreements

Once approved, a license agreement needs to be put in place between the 3rd party and the data controllers. The owners of the data need to ensure to the best of their ability that the data is used ethically and securely. The responsibilities, rights and duties of the database owners and licensee should be clear and agreed at the start of the process. The database owners may have legal departments to write a robust contract, if not then Creative Commons [14] can be used as a template for licence agreements, in that case the owners will need to ensure that this covers all of their requirements.

Depending on the extensiveness of the agreement, it can cover:

- Use: What can the data be used for.
- Content: What information is shared - is it the whole or a subset of the data. What data is available.
- Sharing: can the licensee further share the data to a sub-licensee with permission from the data owner.
- Attribution: The original source of the data must be attributed in any publications.
- Data protection: The data must be securely held. No attempt at identifying personal information of the people imaged.
- Costs: are there any costs for the licensee.
- Termination of agreement and duration of usage: how long can the data be used and what happens at the end of the agreements.
- Commercial usage: can the data be used for commercial purposes?
- What images or data that can be kept or must be deleted at the end of the project.

### 7.3 Processes

A process for accessing the data must be developed. The process will depend upon the nature, size and location of storage for the dataset. Assuming a cloud deployment, a bucket-to-bucket transfer is often a reasonable method. The third-party provisions a bucket in the same cloud provider and onboard a service account to the bucket. The central team then pushes the subset of the dataset to the bucket. This relies on the ability of the central team to work with the 3rd party to define the characteristics and selection criteria for this subset. This can be a long drawn out and complicated process.

## 8 Conclusions

In conclusion, this report outlines the characteristics of frameworks for collecting clinical images and provides guidelines for their safe and effective use. Beyond detailing image and clinical data collection, it also addresses ethical considerations, information governance, and the necessary infrastructure and agreements for data sharing. A key advantage of the primary collection framework discussed is its ability to automate and continuously update datasets, ensuring they remain reflective of current practice. This is essential for training and validating AI tools, as a combination of historical and contemporary data with known outcomes enhances the reliability of assessments. While validations on older datasets provide insights into past conditions, an up-to-date validation dataset ensures models, including AI models, are tested against data they would encounter in real-world deployment. Additionally, alternative collection frameworks that do not rely on full automation are explored. Although a fully automated approach is recommended for large-scale, long-term data collection, semi-automated methods may serve as an initial step, with guidance provided on their implementation.

## 9 Acknowledgments

The authors thank Elizabeth Cooke for reviewing the manuscript and providing insightful suggestions that significantly improved its clarity and readability.

The project 22HLT05 MAIBAI has received funding from the European Partnership on Metrology, co-financed from the European Union's Horizon Europe Research and Innovation Programme and by the Participating States. Funding for the UK partners was provided by Innovate UK under the Horizon Europe Guarantee Extension.

## Appendix 1: Set up and description of dedicated server for collecting images

This Appendix gives more detailed information about a specific automated data collection system: the Royal Surrey NHS Foundation Trust (RSFT) Secure Medical Anonymiser for Research Trials (SMART) box. Examples of site setup guide and checklist; maintenance plan; and user acceptance test are included in the sections below.

### A1.1 SMART Site Setup Guide

### A1.1.1 Introduction

This document is intended to detail the guidelines required to set up a new site and submit images and data to one or many research trials/projects via the semi-automated infrastructure provided by the RSFT's Scientific Computing (SciCom) department. In order to progress the setup process requires a nominated local champion at the collection site to facilitate access to the correct individuals to allow setup to take place.

### A1.1.2 Summary of the Image/Data Collection Process and Setup Requirements

The first principle of the image and data collection system is that all data that is transferred from the clinical sites to the trial endpoint (normally cloud storage) is fully de-identified and that at no point do researchers or third parties have access to client identifiable information. Figure 1 provides a graphical overview of the image and data collection, and storage procedures used.

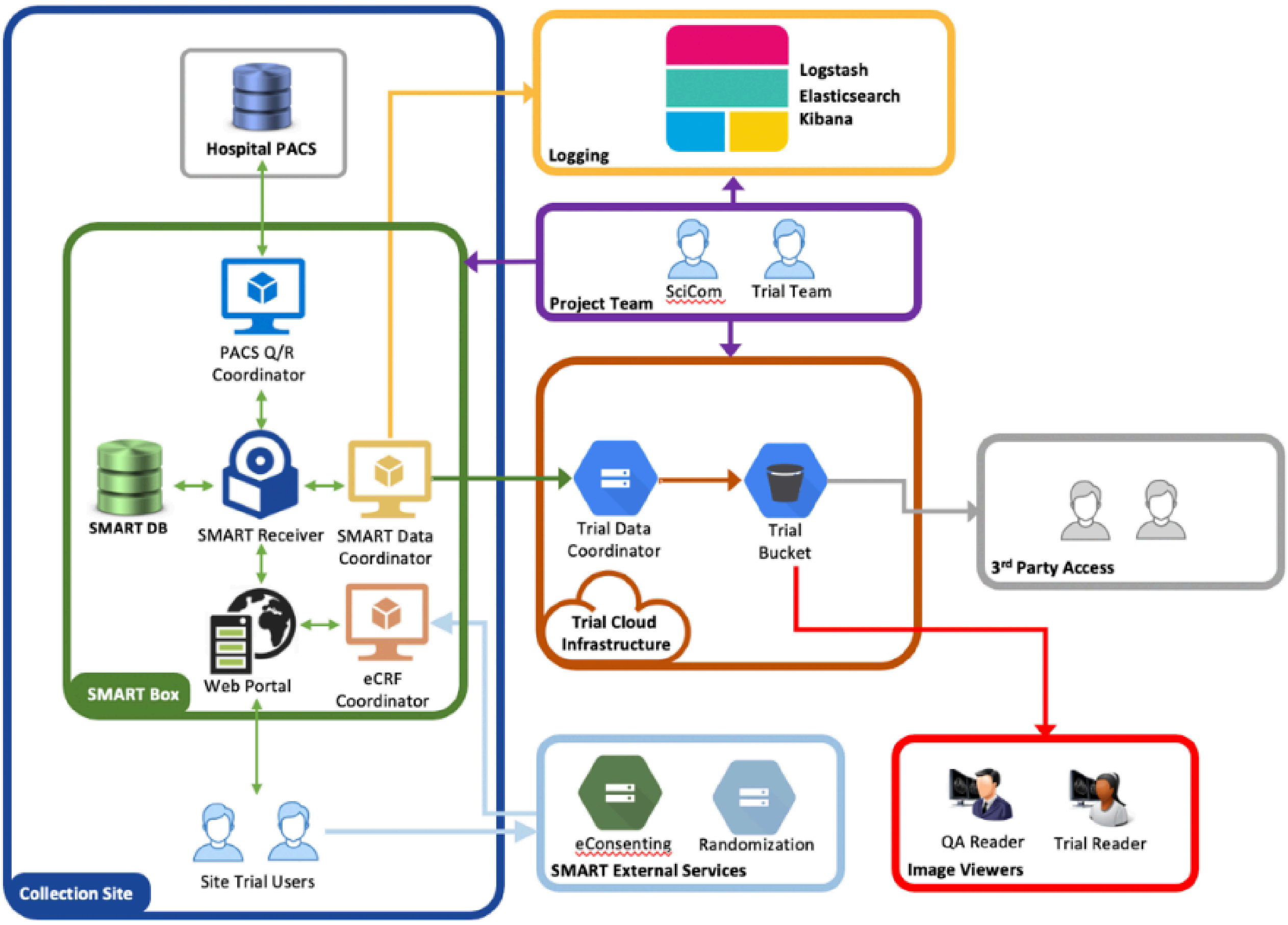

**Figure 3: Graphical overview of the SMART box setup.**

A virtual machine is required at the collection site to act as a collection server (SMART box) which will present both an internal web portal for data management and a DICOM collector for image de-identification. This SMART box has been created, set up and managed by staff at the RSFT and acts as an image/data collection and de-identification manager. The SMART box is configurable and can be used for multiple trials/research projects simultaneously.

The SMART box fulfils three primary roles: Client and data management through an internal web portal, DICOM image collection and de-identification through a DICOM server and data transfer to the trial endpoint. A secure internal web portal provides the functionality to register new clients (providing their NHS number and radiology ID/Hospital number) and to upload a list of clients to submit to the trial along with customisable trial specific data points. Once the clients are registered, the DICOM server is ready to pull images for these clients from the PACS. The SMART box can also be configured to receive the images from a DICOM push if needed.

Depending on the trial, the identification of the clients to be collected may be automated or manual.

If manual, then the submission process is as follows:

1. Login to SMART internal web portal - from here you can view the list of clients that have been uploaded and whether images have been received and submitted to trials
2. Upload the new list of clients to be submitted (NHS number, Hospital Number, other clinical data points)
3. The SMART PACS Q/R Co-ordinator will then automatically pull the corresponding images from the PACS
4. Once set, the SMART box will de-identify the images and data and transfer to the trial endpoint.

***SMART box server requirements at the site***

A virtual machine, ideally running the trusts preferred flavour of Windows Server (however any Operating System (OS) is acceptable) is required at the site. The recommended minimum specifications are: 1 x 3.0GHz 64 Bit Processor (or equivalent), 8 GB RAM and 500 GB Disk space.

The collection site's champion will need to liaise with the Information Governance department, Research & Development and PACS/IT managers.

The PACS/IT manager will be required to facilitate the setup of the VM. The PACS/IT manager will be required to provide a fixed IP (Internet Protocol) address for this server. The following network or firewall exceptions will be required

- The server needs to be able access the internet in order to download updates as well as be able to push data (using Rclone) to the trial endpoint (Ports 443) – specific domains can be provided.
- Server needs to Query/Retrieve images from the PACS. The SMART box will need to be added as a Q/R destination node in PACS.
- SMART box may need to connect to existing clinical databases (e.g., National Breast Screening System (NBSS) via ODBC over port 1972)
- SMART box is required to present an internal web portal via port 8080
- Remote connectivity – The SMART box is able to pull down configuration changes from the cloud, however it is vital that nominated RSFT staff have the ability to remotely administer the SMART box. We are flexible as to which technology/approach that is utilised to facilitate remote connectivity, but must ensure at least one of the following is available:
    - VMWare Horizon
    - Citrix (or alternative)
    - Microsoft Web-based Remote Desktop
    - TeamViewer (or alternative) – Can be a temporary solution
    - Virtual Private Network
- Remote Desktop Protocol (RDP) (port 3389 UDP/TCP (User Datagram Protocol/Transmission Control Protocol)) from specific IP addresses. This is the least desirable as would require firewall changes.

### A1.1.3 Image/Data collection process facilitated by the SMART box

The following section details the operational functionality of the SMART box that will be utilised for the trial collection including Date De-identification, client registration and clinical data upload, image collection and de-identification and data transfer.

***Data de-identification***

Clinical data and images are de-identified at the point of submission/collection. Upon upload, a pseudonym will be assigned to the client. This will be created by a lossful encoding algorithm and trial specific complex salt which will produce an encoded pseudonym that allows the linking of the clinical data with the images. In addition, the NHS/CHI number will also be encrypted using an Advanced Encryption Standard (AES) algorithm and a complex salt to allow linking with national datasets. **No client identifiable information will leave the clinical site.**

### *Client Registration and Clinical Data Upload*

The SMART box presents an internal web portal allowing the upload of a spreadsheet template which can be downloaded from the homepage of the SMART web portal.

Access is granted to the SMART web portal via an administrator account provided during the site setup. This administrative account is able to manage other user accounts with permission to upload data to the SMART box.

When a template is uploaded, the new clients submitted will be automatically registered on the SMART box (separate client registration is not necessary). Only once clients have been registered will the SMART box accept (or fetch) any images for these clients.

*Hospital (internal) ID*

The spreadsheet template contains a column for the NHS number and the Hospital ID. It is vital that both columns are populated and correct in order to ensure that the correct images are assigned to a client. The hospital number (in most cases, unless NHS numbers are used by PACS) is used to link the imaging to the clinical data – hence it is absolutely vital that the hospital number (2nd column) in the spreadsheets are completely accurate and match the ID that Radiology/PACS use as their primary identifier

For example:

- If Radiology use RR12345678 and the hospital use 12345678, then the spreadsheet Hospital Number (2nd column) must use the Radiology representation (RR12345678)
- If Radiology use XX0123456 and the hospital use 123456, then the spreadsheet Hospital Number (2nd Column) must use the Radiology representation (XX0123456)
- If Radiology use NHS numbers as their primary identifier and the hospital use X123456, then the spreadsheet Hospital Number (2nd Column) can use the hospital number (X123456)

*Spreadsheet template*

It is important to note that all trials will require a spreadsheet upload. Trials like OPTIMAM, PROSPECTS, Million Women Study use a more automated approach to identification of trial participants). However, if a trial requires manual identification and upload of clients, then a spreadsheet template is produced for each trial which may contain small or large numbers of columns of data. This is trial specific. All templates will require the NHS/CHI number, Hospital Number as a minimum.

### *Image Collection and De-identification*

Once clients have been registered (as described above), the SMART box is now ready to receive or pull that client's images. There are two options available:

*Pulling images from the SMART box (default)*

Once the patients have been registered, the SMART box can automatically pull the images from the PACS. The images will be received, filtered and de-identified automatically. This is the default and encouraged option.

*Pushing images to the SMART box*

Alternatively, a site might prefer to exercise manual control over studies submitted to a trial. In this instance, once the patients have been registered, the PACS department can be alerted to the list of patients. The PACS staff can then undertake a direct DICOM send to push the images to the SMART box. The images will be received, filtered and de-identified automatically.

### *Data Transfer*

On a nightly basis, any new data and images will be transferred to the trial endpoint using the cloud agnostic tool Rclone. Once transfer has been completed the data and images will be removed from the SMART box and the patient summary table on the SMART box will indicate that the process has been completed. Cloud credentials, providing read and write access to the cloud buckets corresponding to data uploaded by the SMART box only, are provided and stored securely. Uploads are performed using Rclone (using HTTPS (Hypertext Transfer Protocol)) by default, meaning data sent to the cloud is encrypted in transit using TLS (Transport Layer Security). Server-side encrypted is used to ensure that data is encrypted by cloud providers, prior to saving it in its data centres.

## A1.2 SMART Site Setup Checklist

**Site Name and Address:**

| | |
|---|---|
| ☐ Site PACS Lead | Name and Number |
| ☐ IT Lead for the project | Name and Number |

**Server Details**

| | |
|---|---|
| ☐ Server IP Address (Fixed) | Details |
| ☐ Server OS Version | Details |
| ☐ Server disk space assigned (min 500GB) | Details |
| ☐ Server RAM Assigned (min 16GB) | Details |
| ☐ Preferred IT contact details | Details |
| ☐ Preferred Remote Connectivity Option | Details |
| ☐ Remote Connectivity details | Details |

**PACS Details**

| | |
|---|---|
| ☐ PACS Manager Contact Details | Name and Number |
| ☐ Source AEtitle | SMARTDICOMRCV |
| ☐ Source IP (same as Server IP above) | Details |
| ☐ Source Port (can be changed) | 104 |
| ☐ Destination AEtitle (Your PACS) | Details |
| ☐ Destination IP | Details |
| ☐ Destination Port | Details |
| ☐ Destination PACS Role (E.g. Hospital PACS or Screening PACS) | Details |
| ☐ Destination AEtitle 2 (If 2nd PACS e.g. screening) | Details |
| ☐ Destination IP 2 | Details |
| ☐ Destination Port 2 | Details |
| ☐ Destination PACS Role 2 (E.g. Hospital PACS or Screening PACS) | Details |

**NBSS Details**

| | |
|---|---|
| ☐ NBSS Server IP Address | Details |
| ☐ Contact to setup NBSS Account | Details |
| ☐ Contact to reset NBSS Passwords | Details |
| ☐ NBSS Account Details | Username: Password: |

## A1.3 SMART Maintenance Plan

This section defines the maintenance processes for a deployed SMART box including the frequency of updates, monitoring plan and update process. This document is directed at the host institute of the SMART box.

***Definitions***

The following definitions apply throughout this section:

- **SMART box:** The Secure Medical Anonymiser for Research Trials is a collection of software and tools combined to form a configurable DICOM Q/R and Store mode with associated databases and logic to facilitate customisable de-identification profiles for the collection of medical images.

- **Application Administrator:** The Application Administrator is the staff member ultimately responsible for the SMART box.
- **Project Team:** The Project Team is responsible for the creation, delivery and maintenance of the SMART box. It is recommended to have the following roles in the team: Application Administrator, System Administrator and Developer, Project Manager.
- **Research Server:** The Research Server describes the physical or virtual server hosted at the host institute onto which the SMART box is installed and operated from. The host institute is responsible for provision of power, environmental conditioning and for maintenance of Operating System security patches and domain specific rule enforcement.
- **Third Party/Auxiliary Software:** Third Party/Auxiliary Software refers to any applications or software tools that are utilised by the SMART box but are not developed by the project team. For example, the Java Runtime Environment and Apache.

***Connectivity***

The personnel maintaining the SMART box will require ongoing remote access to the Research Server.

The Research Server will need to maintain the following internal and external connectivity rules:

1. Internal connection from Research Server to PACS or vendor neutral archive over DICOM network protocols.
2. Internal connection from Research Server to NBSS server over ODBC (TCP) network protocols.
3. External Connection to bitbucket domain to allow configuration files to be updated.
4. External Connection to Google Cloud Platform buckets to allow data upload.
5. External Connection opened to enable remote connection for the project team.

***Training***

The project team is trained in the installation and maintenance of the SMART box through internal onboarding processes at the RSFT. No training is required or given to the host institute staff for the maintenance of the SMART box.

***Deployment Diagram***

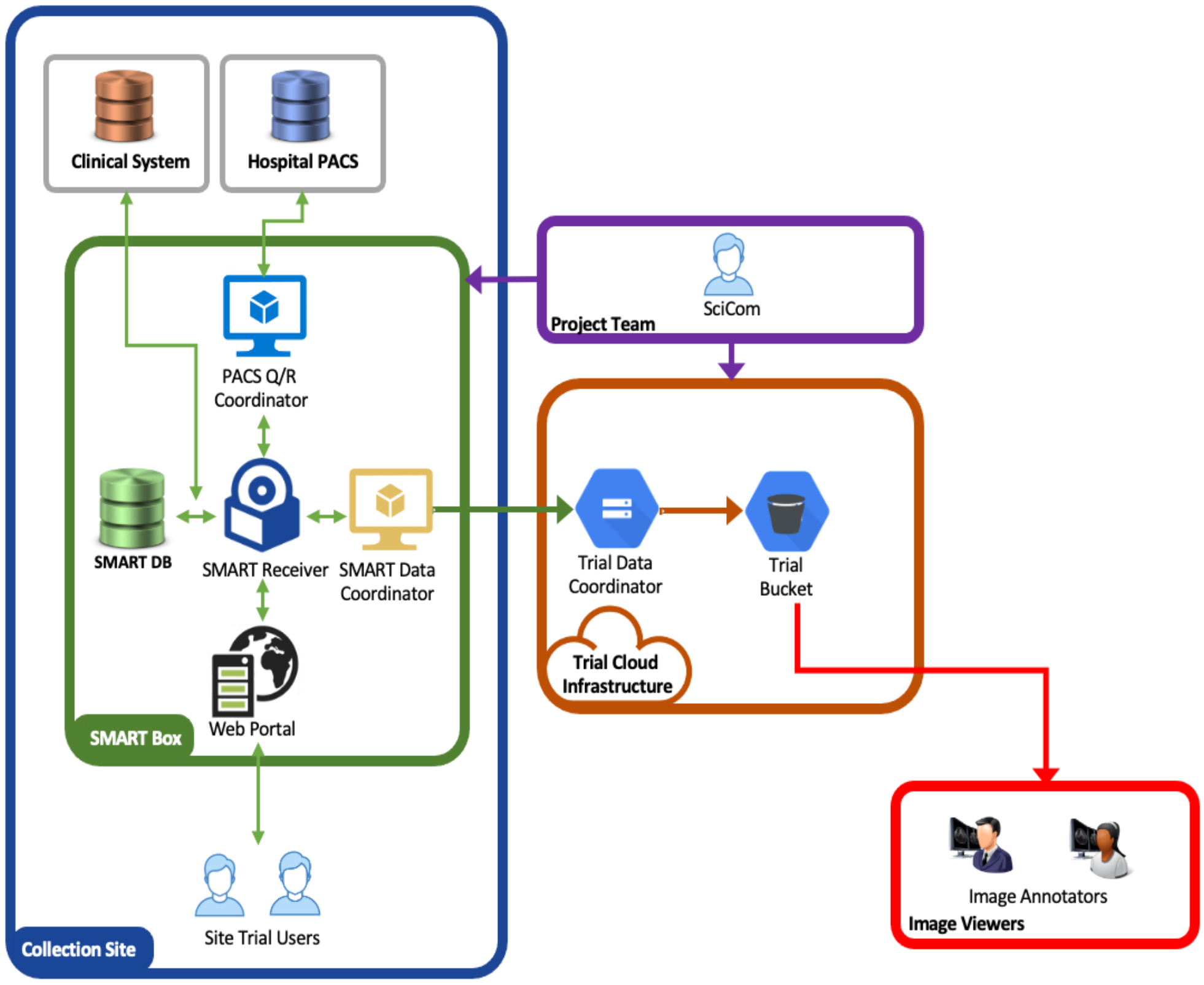

**Figure 4: Graphical overview of the deployment setup at a collection site.**

***Maintenance***

This section covers the proactive and corrective process involved in the maintenance of the SMART box and concerns the maintenance of the Research Server, SMART box, and Third Party/Auxiliary Software.

*Research Server*

The Research Server is the physical or virtual server provided by the host institute onto which the SMART box is installed and operated.

- **Responsible Team:** The host institute IT team
- **Patch Initiation Process/Monitoring:** In line with the host institute's standard operating procedures (SOPs).
- **Release Frequency/Risk Appetite:** In line with the host institute's SOPs.
- **Patch Window:** In line with the host institute's SOPs. The SMART box can be updated and restarted without any notice or Project Team actions.
- **Corrective maintenance Process:** In line with the host institute's SOPs.
- **Communication:** In line with the host institute's SOPs. The Project Team does not need to be alerted. SMART box will recover automatically.

*SMART Box*

The SMART box is a collection of software and tools (developed by the RSFT) combined to form a configurable DICOM Q/R and Store mode with associated databases and logic to facilitate customisable de-identification profiles for the collection of medical images.

- **Responsible Team:** The RSFT project team are responsible for maintenance of SMART boxes.
- **Release Frequency/Risk Appetite:** The RSFT will adopt a medium risk appetite in relation to updates to the SMART box by ensuring that urgent/security patching are undertaken urgently, however, routine version updates which do not contain urgent security patches will only be rolled out when new features are required. This medium risk approach balances the need to ensure that the SMART box is running on a stable platform (known working version) with the need to patch urgent security flaws.
- **Patch Initiation Process/Monitoring:** Patching to SMART boxes will occur when a security flaw is identified during routine testing or when a new feature is added. SMART boxes include a cloud-based monitoring and reporting system which allow us to remotely monitor the version, activity, disk-space etc. of the local system. This monitoring system is reviewed daily and includes email alerts.
- **Patch Window:** Urgent security patches are to be applied at the earliest point after the alert is received. The SMART box can be stopped at any time without notice.
- **Corrective Maintenance Process**
    1. Update the SMART box tool versions and configurations on the cloud/bitbucket.
    2. SMART box will automatically update the various tools.
    3. Undertake the relevant tests using the remote monitoring systems to ensure the new version has not affected the running of SMART box.
- **Communication:** Updates to Third Party/Auxiliary software will be communicated within the Project Team only.

*Third Party/Auxiliary Software*

The Third Party/Auxiliary software currently employed by a SMART box are:

| Software Name | Source (access date 24/6/25) |
|---|---|
| **Java Runtime Environment** | https://www.oracle.com/uk/java/technologies/downloads/ |
| **Apache** | https://www.apachehaus.com/cgi-bin/download.plx |
| **PHP** | https://windows.php.net/download#php-7.3 |
| **OpenSSL** | Bundled with ApacheHaus |
| **MySQL** | https://dev.mysql.com/downloads/mysql/ |
| **Git** | https://git-scm.com/download/win |
| **Rclone** | https://rclone.org/downloads/ |
| **Notepad++** | https://notepad-plus-plus.org/downloads/ |
| **Python** | https://www.python.org/about/ |
| **NuShell** | https://www.nushell.sh/ |

- **Responsible Team:** The RSFT project team are responsible for maintenance of Third Party/Auxiliary software on the Research Server.
- **Release Frequency/Risk Appetite:** The RSFT will adopt a medium risk appetite in relation to Third Party/Auxiliary software by ensuring that urgent/security patching is undertaken urgently, however, routine version updates, which do not contain urgent security patches will not be updated unless an urgent security patch demands. This medium risk approach balances the need to ensure that the SMART box is running on a stable platform (known working version) with the need to patch urgent security flaws.
- **Patch Initiation Process/Monitoring:** The RSFT IT department provides an internal common software security patch alert service to all departments that request to be on the feed. The project team monitors this feed for any security alerts related to the above 3rd party software. If an alert is received, then patching is initiated.
- **Patch Window**: Urgent security patches are to be applied at the earliest point after the alert is received. The SMART box can be stopped at any time without notice.
- **Corrective Maintenance Process**
    1. Stop SMART box services.
    2. Identify the software requiring updating/patching.
    3. Disable/turn-off or delete the service (if applicable).
    4. Download the updated version of patch.
    5. Install or copy to the correct location according to the SMART setup SOP.
    6. Activate the service (if applicable)
    7. Undertake the relevant tests to ensure the new version has not affected the running of SMART box.
- **Communication:** Updates to Third Party/Auxiliary software will be communicated within the Project Team only.

## A1.4 SMART User Acceptance Testing

**SMART Box - User Acceptance Test (UAT)**

| Test ID | Name |
|---|---|
| Test 01 | Browser Compatibility (Desktop) |
| Test 02 | Certificate Test |
| Test 03 | Login Test |
| Test 04 | Change Password Test |
| Test 05 | Logout Test |
| Test 06 | Register Client Test |
| Test 07 | Batch Registration of Client Test |
| Test 08 | Download Data Test |
| Test 09 | Check Clients Test |
| Test 10 | Single Test Case Test |
| Test 11 | Commissioning Batch Results |
| Test 12 | |
| Test 13 | |
| Test 14 | |

| Comments |
|---|
| |

| | Name | Title | Signature | Date |
|---|---|---|---|---|
| Completed By | | | | |
| Accepted By | | | | |

## UAT Test 01 – Browser Compatibility (Desktop)

| **Objective** |
|---|
| 1. To test the website loads and is formatted correctly on the latest versions of the following desktop web-browsers;<br>a. Google Chrome (*this is the highest priority*)<br>b. Mozilla Firefox<br>c. Microsoft Edge (*Internet Explorer is not supported*)<br>d. Apple Safari (*Only on Apple MacOS*)<br>2. Browser 'extensions' may cause the site to break/work incorrectly, and we cannot guarantee the site will work when ad-blockers and other extensions which interfere with Javascript are being used. The site must be used with the native browser settings, without extensions or modified security settings (e.g. Javascript has not been disabled). |

| **Acceptance Requirements** |
|---|
| 1. The website fully loads, with all elements in the correct place<br>2. Navigation links work correctly<br>3. Formatting of the page appears acceptable, with no "out-of-place" elements or layout corruption |

| **Test Method** |
|---|
| 1. Using each of the browsers listed in the objective, load the website and check the layout appears correct based on previous demonstrations |

| **Test Result** | **Pass / Fail** | **Initials** | **Date** |
|---|---|---|---|
| 1. Google Chrome functions as expected | | | |
| 2. Mozilla Firefox functions as expected | | | |
| 3. Microsoft Edge functions as expected (if able to test) | | | |
| 4. Apple Safari functions as expected (if able to test) | | | |

| **Comments** |
|---|
| |

| **Conclusion** | **Pass / Fail** | **Initials** | **Date** |
|---|---|---|---|
| Test 01 – Browser Compatibility (Desktop) | | | |

| | **Name** | **Title** | **Signature** | **Date** |
|---|---|---|---|---|
| Completed By | | | | |
| Accepted By | | | | |

**UAT Test 02 – Certificate Test**

| **Objective** |
|---|
| 1. To test that the correct SSL certificate is utilised |

| **Acceptance Requirements** |
|---|
| 1. Inspection of the SSL certificate details reveals the correct certificate used |

| **Test Method** |
|---|
| 1. Load the site in a browser and then double-click on the shield symbol to investigate the certificate details |

| **Test Result** | **Pass / Fail** | **Initials** | **Date** |
|---|---|---|---|
| 1. Correct certificate presented | | | |

| **Comments** |
|---|
| |

| **Conclusion** | **Pass / Fail** | **Initials** | **Date** |
|---|---|---|---|
| Test 02 – Certificate Test | | | |

| | **Name** | **Title** | **Signature** | **Date** |
|---|---|---|---|---|
| Completed By | | | | |
| Accepted By | | | | |

**UAT Test 03 – Login Test**

| Objective |
|---|
| 1. To test the log in functionality is working correcting |

| Acceptance Requirements |
|---|
| 1. User is able to login with account details provided |

| Test Method |
|---|
| 1. Attempt to login the SMART portal using account details provided |

| Test Result | Pass / Fail | Initials | Date |
|---|---|---|---|
| 1. User is able to login and is presented with the SMART portal homepage | | | |

| Comments |
|---|
| |

| Conclusion | Pass / Fail | Initials | Date |
|---|---|---|---|
| Test 03 – Login Test | | | |

| | Name | Title | Signature | Date |
|---|---|---|---|---|
| Completed By | | | | |
| Accepted By | | | | |

## UAT Test 04 – Change Password Test

| Objective |
|---|
| 1. To confirm that a user can update their password on the portal as required |

| Acceptance Requirements |
|---|
| 1. A valid "Current Password" must be provided in order to change passwords<br>2. The "New Password" must be at least 8 characters in length |

| Test Method |
|---|
| 1. Log into the SMART Portal using the current password<br>2. Select "Change Password" from the menu options on the left-hand side<br>3. Try entering an invalid "Current Password" and blank "New Password"<br>4. Now try entering an invalid "Current Password" and < 8 character "New Password"<br>5. Enter an invalid "Current Password" and a >= 8 character "New Password"<br>6. Finally, enter the valid "Current Password" and a >= 8 character "New Password"<br>7. After changing the password, click "Logout" and then try to login to the SMART Portal using the "New Password" |

| Test Result | Pass / Fail | Initials | Date |
|---|---|---|---|
| 1. Entering an invalid "Current Password" and blank "New Password" should give errors on the form | | | |
| 2. Entering an invalid "Current Password" and password less than 8 characters in length should give errors on the form | | | |
| 3. Entering an invalid "Current Password" and a valid "New Password" (>= 8 characters in length) should give errors on the form | | | |
| 4. Entering a valid "Current Password" and valid "New Password" should allow you to change passwords, and a success message in green will be displayed informing you of the change | | | |
| 5. Log out and back into the SMART Portal, with the new password | | | |

| Comments |
|---|
| |

| Conclusion | Pass / Fail | Initials | Date |
|---|---|---|---|
| Test 04 – Change Password Test | | | |

| | Name | Title | Signature | Date |
|---|---|---|---|---|
| Completed By | | | | |
| Accepted By | | | | |

### UAT Test 05 – Logout Test

| Objective |
|---|
| 1. To check that logging out of the portal is successful, and only logged in accounts can access the SMART Portal controls |

| Acceptance Requirements |
|---|
| 1. The user can log out of the SMART Portal<br>2. When logged out of the SMART Portal it is not possible to access any SMART Portal controls, even using the direct URL addresses |

| Test Method |
|---|
| 1. Start by logging into the SMART Portal using a valid Username and Password<br>2. Confirm that you have valid access to the SMART Portal by clicking "Register Client" – you should be permitted to see the "Register Client Page"<br>3. Now log out of the SMART Portal by clicking the Logout button<br>4. A message should be displayed in the main window informing you that you have successfully logged out of the SMART Portal<br>5. Confirm that the SMART Portal will not allow unauthorized access by manually changing the URL in the address bar, replacing "logout.php" with "registerclient.php" and trying to access that page. You should be redirected back to the index page and forbidden from seeing restricted pages without being logged in |

| Test Result | Pass / Fail | Initials | Date |
|---|---|---|---|
| 1. The user can log out of the SMART Portal | | | |
| 2. When logged out the user cannot access restricted areas of the SMART Portal | | | |

| Comments |
|---|
| |

| Conclusion | Pass / Fail | Initials | Date |
|---|---|---|---|
| Test 05 – Logout Test | | | |

| | Name | Title | Signature | Date |
|---|---|---|---|---|
| Completed By | | | | |
| Accepted By | | | | |

## UAT Test 06 – Register Client Test

| **Objective** | | | |
|---|---|---|---|
| 1. To successfully register a single client using the SMART Portal so that the SMART Collector software will know about the client ahead of collecting the images, so that the images can be successfully processed and linked to the client<br>2. To prevent invalid clients being registered (e.g. with incorrect NHS numbers)<br>3. The confirm that data validation works on the Register Client page | | | |
| **Acceptance Requirements** | | | |
| 1. A test client can be registered on the SMART Portal<br>2. Invalid client data will be rejected by the SMART Portal when entered into the form | | | |
| **Test Method** | | | |
| 1. Login to the SMART Portal using a valid username and password and select "Register Client".<br>2. Select the OPTIMAM trial from the dropdown box – depending on the current configuration of the SMART Collector this might be the only trial you can see, and the dropdown may be disabled.<br>3. The Centre dropdown will be disabled, and should show the correct centre where the test is being run<br>4. Start by entering some invalid text (letters and numbers) in the Primary ID (NHS/CHI number) and click Register Client. It should fail with an error message telling you the NHS number is invalid<br>5. Now enter a non-NHS number that looks like it might be valid (10 digits long and only numbers) – A good example to use is "1234567890" – again this will fail with an error message as this isn't a valid NHS number<br>6. Now enter a valid NHS number – "9999999999" – this is a valid NHS number. The form will still fail because a "Trial Code" has not been provided<br>7. Using that same test NHS number, "9999999999" enter a trial-code, e.g. "UAT-TESTING-01" and try to register again, this time it should be successful, a green success message will be shown at the top of the page<br>8. Try registering a second client with the NHS number "8888888888" and the trial-code "UAT-TESTING-01" again – and error should tell you that no two clients can have the same trial code<br>9. Now try changing the NHS number back to "9999999999" and provide a different trial-code, e.g. "UAT-TESTING-02" – again you'll get an error telling you that this client has already been registered<br>10. Finally, you can test the date enrolment field – this is an optional field, and not required to register a client, however it can be helpful to provide. Try some random text in this box, and then try a valid date. Confirm that appropriate error messages are generated | | | |
| **Test Result** | **Pass / Fail** | **Initials** | **Date** |
| 1. NHS number validation is working – invalid NHS numbers are rejected | | | |
| 2. Trial code validation is working – a trial code must be provided for each client | | | |
| 3. Client duplication is prevent<br>4. ed – clients cannot be registered with the same NHS number and/or trial code | | | |
| 4. Date validation is working | | | |
| **Comments** | | | |
| | | | |
| **Conclusion** | **Pass / Fail** | **Initials** | **Date** |
| Test 06 – Register Client Test | | | |

| | **Name** | **Title** | **Signature** | **Date** |
|---|---|---|---|---|
| Completed By | | | | |
| Accepted By | | | | |

### UAT Test 07 – Batch Registration of Client Test

**Objective**

1. When registering lots of clients into the SMART Portal at the same time it's far easier and quicker to use the batch registration feature, using an Excel spreadsheet

**Acceptance Requirements**

1. Batch registration of clients is successful and errors within the Excel spreadsheet are detected

**Test Method**

1. Login to the SMART Portal and select "Batch Registration" from the left-hand menu
2. On the main page there is a green download link entitled "downloaded here" – use this to download the blank Excel template file
3. Edit this file in Excel and enter a few rows of test data. You want to use a mix of valid and invalid data rows, to confirm that data validation is taking place. For example, you might enter 6 different rows, with data as follows:

| Primary ID (NHS Number) | Secondary ID (Hospital / Radiology Number) | Trial Code | Date Enrolled (Optional) |
|---|---|---|---|
| 1111111111 | Test1 | UAT-TESTING-02 | |
| 2222222222 | Test2 | UAT-TESTING-03 | 44/33/2043 |
| | Test3 | UAT-TESTING-04 | |
| 1234567890 | Test4 | UAT-TESTING-05 | |
| This is not a number | Test5 | UAT-TESTING-06 | |
| 3333333333 | Test6 | | |

4. Save the Excel file with data as provided in the table, and then in the Portal click the "Select Batch Excel File" and choose this saved Excel file – assuming you use the data provided in the table above you will get a message pop up telling you there are 5 errors, and the data cannot be uploaded. This is expected – click ok, and then you will see a red box telling you what the errors are, and where they occurred on the spreadsheet
5. You should have a Row 3, Row 4, Row 5, Row 6 and Row 7 error – with different reasons given (e.g. invalid NHS number, invalid date, missing trial code etc.)
6. Fix the spreadsheet by removing rows 3 through 7, and then try to upload the spreadsheet again, when no errors are present in the spreadsheet you will get a confirmation box asking if you want to proceed, and it will tell you how many clients will be registered. Click ok to continue
7. You'll get a green success message telling you that clients have been uploaded into the database

| Test Result | Pass / Fail | Initials | Date |
|---|---|---|---|
| 1. When a spreadsheet contains errors, these will be displayed to the user in a red box on the page | | | |
| 2. When the spreadsheet contains no errors, the clients will be uploaded into the database with a green box success message displayed | | | |

**Comments**

| Conclusion | Pass / Fail | Initials | Date |
|---|---|---|---|
| Test 07 – Batch Registration of Client Test | | | |

| | Name | Title | Signature | Date |
|---|---|---|---|---|
| Completed By | | | | |
| Accepted By | | | | |

### UAT Test 08 – Download Data Test

**Objective**

1. The SMART Portal provides a comprehensive mechanism to obtain the "ground truth" of all clients/studies/images that have been successful collected and processed by the box
2. This data is provided by an Excel spreadsheet, containing different tabs for Overview, Clients, Studies and Images, which allows a comprehensive exploration of all data that has been processed to date

**Acceptance Requirements**

1. The download data spreadsheet can be generated and downloaded

| Test Method |
| --- |
| 1. Login to the SMART Portal and select “Download Data” from the left-hand menu<br>2. Make sure OPTIMAM is selected if the Trial dropdown box is active (this depends on the user account privileges being used)<br>3. In the export options click all checkboxes (this can be used to restrict the data generated in the spreadsheet, but generally you always want to see all data)<br>4. Open the Excel file and confirm it contains data – depending on when the user acceptance testing is done, there may or may not be much/any data in the spreadsheet, as it depends on whether the collector has already started receiving and processing images<br>5. As long as the Excel sheet is generated, the test is passed |

| Test Result | Pass / Fail | Initials | Date |
| --- | --- | --- | --- |
| 1. An Excel spreadsheet is successfully downloaded with the “Download Data” feature, regardless of whether the Excel sheet is populated with contents or not, as this depends on whether the collector has already started receiving images | | | |

| Comments |
| --- |
| |

| Conclusion | Pass / Fail | Initials | Date |
| --- | --- | --- | --- |
| Test 08 – Download Data Test | | | |

| | Name | Title | Signature | Date |
| --- | --- | --- | --- | --- |
| Completed By | | | | |
| Accepted By | | | | |

### UAT Test 09 – Check Clients Test

| Objective |
|---|
| 1. The SMART Portal can be used to quickly check the registration status of one or more clients<br>2. An Excel spreadsheet will be generated containing one row per client, with some information about their collection status |

| Acceptance Requirements |
|---|
| 1. The “Check Clients” page works as expected |

| Test Method |
|---|
| 1. Login to the SMART Portal and select “Check Clients” from the left-hand menu<br>2. Make sure OPTIMAM is selected in the Trial, assuming the dropdown box is enabled<br>3. Enter one more client search terms in the box – these can be separated by commas, spaces and/or new lines, as desired – the search can be on either the primary (NHS Number) or secondary (Hospital/Radiology Number)<br>4. We’ll use some of the test data from previous tests, enter the following into the text box:<br>1111111111<br>3333333333<br>THIS_IS_NOT_A_NUMBER<br>9999999999<br>5. Check an Excel sheet is downloaded – assuming the valid NHS number, “9999999999”, was used in UAT test 06, you should see this reflected in the Excel sheet. |

| Test Result | Pass / Fail | Initials | Date |
|---|---|---|---|
| 1. An Excel spreadsheet is successfully downloaded with the “Check Clients” feature. The four test inputs should be shown in the sheet, and one of them, “9999999999”, should be shown as being registered, assuming this same NHS number was used in UAT test 06 | | | |

| Comments |
|---|
| |

| Conclusion | Pass / Fail | Initials | Date |
|---|---|---|---|
| Test 09 – Check Clients Test | | | |

| | Name | Title | Signature | Date |
|---|---|---|---|---|
| Completed By | | | | |
| Accepted By | | | | |

## Appendix 2: Documents of the Data Committees

Appendices A2.1 and A2.2 are example cases from the collection of mammography images for the OPTIMAM database by the Royal Surrey.

### A2.1: DATABASE STEERING COMMITTEE

| **1. Description** | |
|---|---|
| Definition | The Database collects NHS Breast Screening Programme (NHSBSP) screening images from multiple breast screening centres across the UK and has been created to serve as a large repository of de-identified breast images to support research and development involving medical imaging. |
| Outline of scope of Charter | The purpose of this document is to describe the membership, terms of reference, roles, responsibilities, authority, decision-making and relationships of the Database Steering Committee (DSC) for the project, including the timing of meetings, frequency and format of meetings and the relationship with Annual meetings. |
| **2. Roles and responsibilities** | |
| A broad statement of the aims of the DSC | To advise on the development and use of the Database. |
| Specific roles of DSC members | The specific roles of the members of the DSC are to:<br>● attend DSC meetings and advise on availability for future DSC meetings<br>● provide clinical or other expert guidance, where applicable<br>● maintain confidentiality of information that is not in the public domain<br>● respond to correspondence and any questions in a timely fashion<br>● provide responses to any issues or concerns raised to the DSC<br>● consider the statistical validity of datasets and collection, where applicable<br>● contribute to the interpretation and writing of any reports resulting from the DSC activities |
| **3. Before or early in participation in the DSC** | |
| Whether members of the DSC will have a contract | DSC members will not formally sign a contract. They should formally register their agreement to join the group by confirming (1) that they agree to be on the DSC and (2) that they agree with the contents of this Charter. Members should complete and return the form in Annexe 2. |
| **4. Composition** | |
| Membership and size of the DSC | The members of the DSC can be found in Annexe 1.<br>Members of the DSC should disclose potential competing interests. |
| The Chair, how they are chosen and the Chair's role (Likewise, if relevant, the vice-Chairman) | The Chair of the DSC will be the Project Lead of the project. They should also have experience of Chairing meetings and should be able to facilitate and summarise discussions. (See Organisation of DSC Meetings)<br>The vice-Chair will assume the responsibilities of the Chair when the Chair is not available. |
| The role of Patient and Public Involvement (PPI) representatives | The DSC will have at least one PPI representative, replacement members (if required) will be arranged through the Independent Cancer Patient's Voice in the first instance. The PPI representatives will provide an important perspective on all aspects of the activities of the DSC. The chair will provide additional support for PPI members. |

| **5. Organisation of meetings** | |
|---|---|
| Expected frequency of DSC meetings | The DSC will meet via teleconference for two Steering Committee meetings per year in addition to one face-to-face Annual meeting. Appropriate colleagues from outside the DSC will be invited to attend the Annual Meeting to hear about the work in the project. |
| Whether meetings will be face-to-face or by teleconference | Meetings will be a mixture of teleconference and face-to-face, depending on the needs of the database. Meetings will be organised by Royal Surrey staff. |
| When the DSC is quorate | When there is at least one representative from both clinical and technical areas. |
| Can DSC members who cannot attend the meeting input | The agenda will be circulated before the meeting; DSC members who will not be able to attend the meeting may pass comments to the DSC Chair and Royal Surrey (RS) for consideration during the discussions. |
| **6. Reporting** | |
| Whether minutes of the meeting be made and, if so, by whom and where they will be kept | A short summary of each meeting with any action points clearly marked will be made by the RS team. Initial drafts will be made by Royal Surrey. The DSC Chair will sign off these notes and a copy will be circulated to the DSC. |

## A2.2: DATABASE ACCESS COMMITTEE

| **1. Description** | |
|---|---|
| Description | The Database collects NHS Breast Screening Programme (NHSBSP) screening images from multiple breast screening centres across the UK and has been created to serve as a large repository of de-identified breast images to support research and development involving medical imaging. |
| Outline of scope of Charter | The purpose of this document is to describe the membership, terms of reference, roles, responsibilities, authority, decision-making and relationships of the Database Access Requests Sub-committee for the project, including the reviewing of and decision-making processes for applications for access to the database. |
| **2. Roles and responsibilities** | |
| A broad statement of the aims of the Data Access Committee (DAC) | To advise on the technical or clinical merit (pursuant to specialty), of the projects proposed by applicants in which they wish to use the data from the Database. |
| Specific roles of DAC members | The specific roles of the members of the DA are to:<br>● provide technical, clinical or other expert guidance, where applicable<br>● maintain confidentiality of information that is not in the public domain<br>● respond to correspondence and any questions in a timely fashion<br>● provide responses to any issues or concerns raised to the DAC |
| **3. Before or early in participation in DAC** | |
| Whether members of the DAC will have a contract | DAC members will not formally sign a contract. They should formally register their agreement to join the group by confirming (1) that they agree to be on DAC and (2) that they agree with the contents of this Charter. Members should complete and return the form in A.2.2.2. |
| **4. Composition** | |
| Membership and size of DAC | The members of DAC can be found in A.2.2.1.<br>Members of DAC should disclose potential competing interests (A.2.2.2). |
| The Chair, how they are chosen and the Chair's role | The Chair of DAC will be the Project Lead of the DAC project.<br>The vice-Chair/s will assume the responsibilities of the Chair when the Chair is not available. |
| **5. Organisation of review of Applications for Access to the Database** | |
| Expected frequency of Database applications for access to review | This will vary upon demand, and applications for access to the Database may be more frequent when the database has been referenced in a publication for example. |
| When the DAC is quorate | When there is at least one representative from both clinical and technical areas. |
| **6. Reporting** | |
| Whether reviews will be kept and, if so, by whom and where they will be kept | The Royal Surrey will keep reviews that have been provided by DAC members for applications for access to the Database. They will be used to inform a final decision in approving or not approving an application, by the chair or vice-chair/s. |

### A2.2.1: DAC Members

| **NAME** | **ROLE** |
|---|---|
| **__________________ Organisation Name** | |
| | |
| | |
| | |

### A2.2.2: Agreement and potential competing interests' form

Please complete the following document and return to the CI.

(Please initial box to agree)

☐ I have read and understood the Database Sub-Committee Charter version **1 dated 29-Jan-2025**

☐ I agree to join the Database Access Requests Sub-committee

☐ I agree to treat all sensitive data and discussions confidentially

The avoidance of any perception that members of a Sub-committee may be biased in some fashion is important for the credibility of the decisions made by DAC and for the integrity of the database.

Potential competing interests should be disclosed via the coordinating centre, Royal Surrey. In many cases simple disclosure up front should be sufficient. Otherwise, the (potential) DAC member should remove the conflict or stop participating in DAC. **Table 1** lists potential competing interests.

☐ **No,** I have no potential competing interests to declare

☐ **Yes,** I have potential competing interests to declare (please detail below)

Please provide details of any potential competing interests:

______________________________________________________________

______________________________________________________________

______________________________________________________________

Name: ____________________________

Signed: ___________________________ Date: ______________

**Table 1: Potential competing interests**

| |
|---|
| ● Employment by any organisation with a financial interest in or financial conflict with the database<br>● Funding source – financial support by any organisation with a financial interest in or financial conflict with the database<br>● Personal financial interests – financial interests such as stocks and shares in companies with a financial interest in or financial conflict with the database; consultation fees or other remuneration from organisations with a financial interest in or financial conflict with the database; patents or patent applications which may be affected by the database<br>● Membership of organisations with a financial interest in or financial conflict with the database |

## A2.3: Guidelines for reviewing the data access requests for the database

### A.2.3.1 Data Access Committee

Access requests will be assessed by a committee of experts including:

- 1 or **more** clinical advisors
- 1 or **more** technical scientific advisors
- An administrator to manage the access requests
- A chairperson (this could be any of the advisors above)

### A2.3.2 Review process

Access requests are reviewed according to the following procedure.

- The administrator applies a preliminary screening in order to:
    - Check that all the relevant information has been submitted. If only a few pieces of information are missing, the administrator gets in touch with the applicant to request them.
    - Data access requests that are very incomplete are declined.
- The administrator forwards the request for data access to the Data Access Committee if it passes the preliminary screening.
- The advisors express their availability to review a submission and declare any conflict of interest with respect to the applicant. If an advisor has a conflict of interest with respect to the applicant and/or the proposed project, they cannot evaluate the proposal and the administrator selects another expert to take part in this first stage of the review. All data access requests must be evaluated by at least one clinical and one technical scientific advisor.
- If the advisors request more information, the administrator contacts the applicant for the information.
    - If the additional information clarifies the issue, the application is approved
    - If it is not initially clear, the application goes back to the advisors to advise if the additional information clarifies the issue
- If both or all advisors approve the application, the data access request is approved.
- If either advisor declines the application and one approves, the application is reviewed by the chair or a nominated advisor in their absence, to come to a consensus decision.
- If both or all advisors decline the application, the data access request is rejected.

The administrator notifies the outcomes of the review process to the applicant.

The chairperson handles appeals to committee decisions. Declined access requests can only be resubmitted if the applicants have introduced major changes or revisions to their proposition. In most cases, the decision to allow resubmissions belongs to the chairperson.

### A2.3.3 Reviewer Forms

Before they engage with the process for reviewing data access requests, all advisors must sign the Database Access Requests Charter.

### A2.3.4 Review Criteria

All approved data access requests must satisfy the criteria set out below. Reviewers will record their assessment of each proposal by filling in the Reviewer Form.

Reviewers will seek reasons for approving data access requests as opposed to declining them. The motivation for this guideline is that the Database aims to be a resource for accelerating research not hindering it.

Reviewers will evaluate data access requests on their individual merits without attempting to compare or prioritise submissions. The motivation for this guideline is that evaluating the relative strengths of different submissions would make the review process exceedingly slow.

### A2.3.5 Scientific merit

Reviewers will evaluate whether the assumptions underlying the project are based on scientific evidence. In addition, reviewers should assess whether the project:

- Is of scientific benefit
- Has a reproducible methodology
- Has clear and measurable objectives

### A2.3.6 Technical feasibility

Reviewers will evaluate whether the data analysis and modelling techniques proposed in the submission are likely to yield the expected results. The following factors should also be considered as indicative of a good submission:

- Similar modelling approaches have proven successful in a related context
- Appropriate choice of software libraries
- Adequacy of the computing and data-storage infrastructure

### A2.3.7 Ability to deliver the work, and track record

Proposals that include realistic execution plans will be regarded positively. Reviewers will check that the project team composition matches the proposed work in terms of expertise and headcount. The reputation of the applicants will also be assessed to ensure that only trustworthy institutions and companies are granted access to the data. Elements that may result in an institution or company being deemed not reputable include lack of transparent information about them on the Internet, and any evidence of unethical or criminal behaviour. Finally, the evaluation will include the track record of the applicant and their team as evidenced by, for example:

- Academic publications in the field of interest, or in related fields
- Development of software solutions of high standard

### A2.3.8 Reasonable evidence that access to the data can benefit patients

Reviewers will assess whether the proposed project is likely to result in software solutions and research outcomes that support staff and/or benefit patients. Innovations that justify access to the data will, for example:

- Reduce the workload on staff
- Improve medical care of patients
- Lead to more efficient logistics of personalised follow up care
- Optimise hospital resources
- Lower cost

### A2.3.9 IT security

Data from the database can only be accessed by named individuals in an approved data access request. These users must not share or disclose database data to anyone else and must under all circumstances store database data in a safe location. Technical reviewers will evaluate the security of the data storage and computing infrastructure proposed in the submission, taking into consideration:

- Application security and access control
- Security of data at rest and in transit
- Logging and auditing
- Preparedness and incident response

## Appendix 3: Details of DICOM tag handling during de-identification from both screening and non-screening settings

Primary: De-identification taking place at the source site.
Secondary: Further de-identification (occurring on the OMI-DB cloud) before 3rd party sharing.

| Tag | VR | Description | Data Replacement | Notes | Stage Implemented |
|---|---|---|---|---|---|
| Private Tags | ? | Private Tags | Removed | | Primary |
| 0032,1021 | AE | Scheduled Study Location AE Title | Data anonymised | | Primary |
| 0040,0001 | AE | Scheduled Station AE Title | Data anonymised | | Primary |
| 0040,0241 | AE | Performed Station AE Title | Data anonymised | | Primary |
| 0010,0040 | CS | Patient's Sex | | Blank string inserted | Primary |
| 0010,21A0 | CS | Smoking Status | | Blank string inserted | Primary |
| 0010,2203 | CS | Patient's Sex Neutered | | Blank string inserted | Primary |
| 0012,0062 | CS | Patient Identity Removed | | Blank string inserted | Primary |
| 0020,3401 | CS | Modifying Device ID | | Blank string inserted | Primary |
| 0008,0012 | DA | Instance Creation Date | Dates offset or set to 19000101 if null or '' if empty | Dates are offset by a random number of days | Primary |
| 0008,0020 | DA | Study Date | Dates offset or set to 19000101 if null or '' if empty | Dates are offset by a random number of days | Primary |
| 0008,0021 | DA | Series Date | Dates offset or set to 19000101 if null or '' if empty | Dates are offset by a random number of days | Primary |
| 0008,0022 | DA | Acquisition Date | Dates offset or set to 19000101 if null or '' if empty | Dates are offset by a random number of days | Primary |
| 0008,0023 | DA | Content Date | Dates offset or set to 19000101 if null or '' if empty | Dates are offset by a random number of days | Primary |
| 0008,0024 | DA | Overlay Date | Dates offset or set to 19000101 if null or '' if empty | Dates are offset by a random number of days | Primary |
| 0008,0025 | DA | Curve Date | Dates offset or set to 19000101 if null or '' if empty | Dates are offset by a random number of days | Primary |
| 0010,0030 | DA | Patient's Birth Date | Set to 0101YYYY (where YYYY is actual YOB) or 01010101 | | Primary |
| 0010,21D0 | DA | Last Menstrual Date | Dates offset or set to 19000101 if null or '' if empty | Dates are offset by a random number of days | Primary |
| 0018,700C | DA | Date of Last Detector Calibration | Dates offset or set to 19000101 if null or '' if empty | | Primary |
| 0038,0020 | DA | Admitting Date | Dates offset or set to 19000101 if null or '' if empty | | Primary |
| 0040,0002 | DA | Scheduled Procedure Step Start Date | Dates offset or set to 19000101 if null or '' if empty | Dates are offset by a random number of days | Primary |
| 0040,0004 | DA | Scheduled Procedure Step End Date | Dates offset or set to 19000101 if null or '' if empty | Dates are offset by a random number of days | Primary |
| 0040,0244 | DA | Performed Procedure Step Start Date | Dates offset or set to 19000101 if null or '' if empty | Dates are offset by a random number of days | Primary |
| 0010,1020 | DS | Patient's Size | 0 | | Primary |
| 0010,1030 | DS | Patient's Weight | 0 | | Primary |
| 0008,002A | DT | Acquisition DateTime | 19000101000000 | | Primary |
| 0008,0080 | LO | Institution Name | Data anonymised | | Primary |
| 0008,1030 | LO | Study Description | Data anonymised | | Primary |
| 0008,103E | LO | Series Description | Data anonymised | | Primary |
| 0008,1040 | LO | Institutional Department Name | Data anonymised | | Primary |
| 0008,1080 | LO | Admitting Diagnoses Description | Data anonymised | | Primary |
| 0010,0020 | LO | Patient ID | | Replaced with pseudonym (mapping stored in sqlite db) | Primary & Secondary |
| 0010,0021 | LO | Issuer of Patient ID | Data anonymised | | Primary |
| 0010,1000 | LO | Other Patient IDs | Data anonymised | | Primary |
| 0010,1040 | LO | Patient's Address | Data anonymised | | Primary |
| 0010,1050 | LO | Insurance Plan Identification | Data anonymised | | Primary |
| 0010,1080 | LO | Military Rank | Data anonymised | | Primary |
| 0010,1081 | LO | Branch of Service | Data anonymised | | Primary |
| 0010,1090 | LO | Medical Record Locator | Data anonymised | | Primary |
| 0010,2000 | LO | Medical Alerts | Data anonymised | | Primary |
| 0010,2110 | LO | Allergies | Data anonymised | | Primary |
| 0010,2150 | LO | Country of Residence | Data anonymised | | Primary |
| 0010,2152 | LO | Region of Residence | Data anonymised | | Primary |
| 0010,21F0 | LO | Patient's Religious Preference | Data anonymised | | Primary |
| 0010,2299 | LO | Responsible Organization | Data anonymised | | Primary |
| 0012,0010 | LO | Clinical Trial Sponsor Name | Data anonymised | | Primary |
| 0012,0020 | LO | Clinical Trial Protocol ID | Data anonymised | | Primary |
| 0012,0021 | LO | Clinical Trial Protocol Name | Data anonymised | | Primary |
| 0012,0030 | LO | Clinical Trial Site ID | Data anonymised | | Primary |
| 0012,0031 | LO | Clinical Trial Site Name | Data anonymised | | Primary |
| 0012,0040 | LO | Clinical Trial Subject ID | Data anonymised | | Primary |
| 0012,0042 | LO | Clinical Trial Subject Reading ID | Data anonymised | | Primary |
| 0012,0050 | LO | Clinical Trial Time Point ID | Data anonymised | | Primary |
| 0012,0051 | LO | Clinical Trial Time Point Description | Data anonymised | | Primary |
| 0012,0060 | LO | Clinical Trial Coordinating Centre Name | Data anonymised | | Primary |
| 0018,0010 | LO | Contrast/Bolus Agent | Data anonymised | | Primary |
| 0018,1000 | LO | Device Serial Number | Data anonymised | | Primary |
| 0018,1004 | LO | Plate ID | Data anonymised | | Primary |
| 0018,1005 | LO | Generator ID | Data anonymised | | Primary |
| 0018,1007 | LO | Cassette ID | Data anonymised | | Primary |
| 0018,1008 | LO | Gantry ID | Data anonymised | | Primary |
| 0018,1020 | LO | Software Version(s) | Data anonymised | | Primary |
| 0018,1030 | LO | Protocol Name | Data anonymised | | Primary |
| 0018,1400 | LO | Acquisition Device Processing Description | Data anonymised | | Primary |
| 0020,3404 | LO | Modifying Device Manufacturer | Data anonymised | | Primary |
| 0020,3406 | LO | Modified Image Description | Data anonymised | | Primary |
| 0032,0012 | LO | Study ID Issuer | Data anonymised | | Primary |
| 0032,1020 | LO | Scheduled Study Location | Data anonymised | | Primary |
| 0032,1030 | LO | Reason for Study | Data anonymised | | Primary |
| 0032,1033 | LO | Requesting Service | Data anonymised | | Primary |
| 0032,1060 | LO | Requested Procedure Description | Data anonymised | | Primary |
| 0032,1070 | LO | Requested Contrast Agent | Data anonymised | | Primary |
| 0038,0010 | LO | Admission ID | Data anonymised | | Primary |
| 0038,0011 | LO | Issuer of Admission ID | Data anonymised | | Primary |
| 0038,001E | LO | Scheduled Patient Institution Residence | Data anonymised | | Primary |
| 0038,0040 | LO | Discharge Diagnosis Description | Data anonymised | | Primary |
| 0038,0050 | LO | Special Needs | Data anonymised | | Primary |
| 0038,0060 | LO | Service Episode ID | Data anonymised | | Primary |
| 0038,0061 | LO | Issuer of Service Episode ID | Data anonymised | | Primary |
| 0038,0062 | LO | Service Episode Description | Data anonymised | | Primary |
| 0038,0300 | LO | Current Patient Location | Data anonymised | | Primary |
| 0038,0400 | LO | Patient's Institution Residence | Data anonymised | | Primary |
| 0038,0500 | LO | Patient State | Data anonymised | | Primary |
| 0040,0007 | LO | Scheduled Procedure Step Description | Data anonymised | | Primary |
| 0040,0012 | LO | Pre-Medication | Data anonymised | | Primary |
| 0040,0254 | LO | Performed Procedure Step Description | Data anonymised | | Primary |
| 0040,1004 | LO | Patient Transport Arrangements | Data anonymised | | Primary |
| 0040,1005 | LO | Requested Procedure Location | Data anonymised | | Primary |
| 0040,1103 | LO | Person's Telephone Numbers | Data anonymised | | Primary |
| 0040,2001 | LO | Reason for the Imaging Service Request | Data anonymised | | Primary |
| 0040,2016 | LO | Placer Order Number / Imaging Service Request | Data anonymised | | Primary |
| 0040,2017 | LO | Filler Order Number / Imaging Service Request | Data anonymised | | Primary |
| 0040,3001 | LO | Confidentiality Constraint on Patient Data Description | Data anonymised | | Primary |
| 0040,4036 | LO | Human Performer's Organization | Data anonymised | | Primary |
| 0040,A027 | LO | Verifying Organization | Data anonymised | | Primary |
| 0088,0904 | LO | Topic Title | Data anonymised | | Primary |
| 0088,0910 | LO | Topic Author | Data anonymised | | Primary |
| 0088,0912 | LO | Topic Keywords | Data anonymised | | Primary |
| 2030,0020 | LO | Text String | Data anonymised | | Primary |
| 4008,0042 | LO | Results ID Issuer | Data anonymised | | Primary |
| 4008,011A | LO | Distribution Address | Data anonymised | | Primary |
| 4008,0202 | LO | Interpretation ID Issuer | Data anonymised | | Primary |
| 0008,4000 | LT | Identifying Comments | Data anonymised | | Primary |
| 0010,21B0 | LT | Additional Patient History | Data anonymised | | Primary |
| 0010,4000 | LT | Patient Comments | Data anonymised | | Primary |
| 0018,4000 | LT | Acquisition Comments | Data anonymised | | Primary |
| 0018,9424 | LT | Acquisition Protocol Description | Data anonymised | | Primary |
| 0020,4000 | LT | Image Comments | Data anonymised | | Primary |
| 0020,9158 | LT | Frame Comments | Data anonymised | | Primary |
| 0028,4000 | LT | Image Presentation Comments | Data anonymised | | Primary |
| 0032,4000 | LT | Study Comments | Data anonymised | | Primary |
| 0038,4000 | LT | Visit Comments | Data anonymised | | Primary |
| 0040,1400 | LT | Requested Procedure Comments | Data anonymised | | Primary |
| 0040,2400 | LT | Imaging Service Request Comments | Data anonymised | | Primary |
| 4000,0010 | LT | Arbitrary | Data anonymised | | Primary |
| 4000,4000 | LT | Text Comments | Data anonymised | | Primary |
| 4008,0115 | LT | Interpretation Diagnosis Description | Data anonymised | | Primary |
| 60XX,4000 | LT | Overlay Comments | Data anonymised | | Primary |
| 0400,0404 | OB | MAC | 0 | Pair of bytes inserted (2x 8-bit zeros) | Primary |
| 50XX,XXXX | OB | Curve Data | 0 | Pair of bytes inserted (2x 8-bit zeros) | Primary |
| 60XX,3000 | OB | Overlay Data | 0 | Pair of bytes inserted (2x 8-bit zeros) | Primary |
| FFFC,FFFC | OB | Data Set Trailing Padding | 0 | Pair of bytes inserted (2x 8-bit zeros) | Primary |
| 0008,0090 | PN | Referring Physician's Name | Data anonymised^^^^ | | Primary |
| 0010,0010 | PN | Patient Name | | Replaced with pseudonym (mapping stored in sqlite db) | Primary & Secondary |
| 0008,1048 | PN | Physician(s) of Record | Data anonymised^^^^ | | Primary |
| 0008,1050 | PN | Performing Physician's Name | Data anonymised^^^^ | | Primary |
| 0008,1060 | PN | Name of Physician(s) Reading Study | Data anonymised^^^^ | | Primary |
| 0008,1070 | PN | Operators' Name | Data anonymised^^^^ | | Primary |
| 0010,1001 | PN | Other Patient Names | Data anonymised^^^^ | | Primary |
| 0010,1005 | PN | Patient's Birth Name | Data anonymised^^^^ | | Primary |
| 0010,1060 | PN | Patient's Mother's Birth Name | Data anonymised^^^^ | | Primary |
| 0010,2297 | PN | Responsible Person | Data anonymised^^^^ | | Primary |
| 0032,1032 | PN | Requesting Physician | Data anonymised^^^^ | | Primary |
| 0040,0006 | PN | Scheduled Performing Physician's Name | Data anonymised^^^^ | | Primary |
| 0040,1010 | PN | Names of Intended Recipients of Results | Data anonymised^^^^ | | Primary |
| 0040,2008 | PN | Order Entered By | Data anonymised^^^^ | | Primary |
| 0040,4037 | PN | Human Performer's Name | Data anonymised^^^^ | | Primary |
| 0040,A075 | PN | Verifying Observer Name | Data anonymised^^^^ | | Primary |
| 0040,A123 | PN | Person Name | Data anonymised^^^^ | | Primary |
| 0070,0084 | PN | Content Creator's Name | Data anonymised^^^^ | | Primary |
| 300E,0008 | PN | Reviewer Name | Data anonymised^^^^ | | Primary |
| 4008,0102 | PN | Interpretation Recorder | Data anonymised^^^^ | | Primary |
| 4008,010A | PN | Interpretation Transcriber | Data anonymised^^^^ | | Primary |
| 4008,010C | PN | Interpretation Author | Data anonymised^^^^ | | Primary |
| 4008,0114 | PN | Physician Approving Interpretation | Data anonymised^^^^ | | Primary |
| 4008,0119 | PN | Distribution Name | Data anonymised^^^^ | | Primary |
| 0008,0050 | SH | Accession Number | Data anonymised | | Primary |
| 0008,0094 | SH | Referring Physician's Telephone Numbers | Data anonymised | | Primary |
| 0008,0201 | SH | Time zone Offset From UTC | Data anonymised | | Primary |
| 0008,1010 | SH | Station Name | Data anonymised | | Primary |
| 0010,2154 | SH | Patient's Telephone Numbers | Data anonymised | | Primary |
| 0010,2160 | SH | Ethnic Group | Data anonymised | | Primary |
| 0010,2180 | SH | Occupation | Data anonymised | | Primary |
| 0018,700A | SH | Detector ID | Data anonymised | | Primary |
| 0020,0010 | SH | Study ID | Data anonymised | | Primary |
| 0040,0010 | SH | Scheduled Station Name | Data anonymised | | Primary |
| 0040,0011 | SH | Scheduled Procedure Step Location | Data anonymised | | Primary |
| 0040,0242 | SH | Performed Station Name | Data anonymised | | Primary |
| 0040,0243 | SH | Performed Location | Data anonymised | | Primary |
| 0040,0253 | SH | Performed Procedure Step ID | Data anonymised | | Primary |
| 0040,1001 | SH | Requested Procedure ID | Data anonymised | | Primary |
| 0040,2009 | SH | Order Enterer's Location | Data anonymised | | Primary |
| 0040,2010 | SH | Order Callback Phone Number | Data anonymised | | Primary |
| 0008,0082 | SQ | Institution Code Sequence | | All sequence and sub-sequence element cleared according to VRs | Primary |
| 0008,0096 | SQ | Referring Physician Identification Sequence | | All sequence and sub-sequence element cleared according to VRs | Primary |
| 0008,1049 | SQ | Physician(s) of Record Identification Sequence | | All sequence and sub-sequence element cleared according to VRs | Primary |
| 0008,1052 | SQ | Performing Physician Identification Sequence | | All sequence and sub-sequence element cleared according to VRs | Primary |
| 0008,1062 | SQ | Physician(s) Reading Study Identification Sequence | | All sequence and sub-sequence element cleared according to VRs | Primary |
| 0008,1072 | SQ | Operator Identification Sequence | | All sequence and sub-sequence element cleared according to VRs | Primary |
| 0008,1084 | SQ | Admitting Diagnoses Code Sequence | | All sequence and sub-sequence element cleared according to VRs | Primary |
| 0008,1110 | SQ | Referenced Study Sequence | | All sequence and sub-sequence element cleared according to VRs | Primary |
| 0008,1111 | SQ | Referenced Performed Procedure Step Sequence | | All sequence and sub-sequence element cleared according to VRs | Primary |
| 0008,1120 | SQ | Referenced Patient Sequence | | All sequence and sub-sequence element cleared according to VRs | Primary |
| 0008,1140 | SQ | Referenced Image Sequence | | All sequence and sub-sequence element cleared according to VRs | Primary |
| 0008,2112 | SQ | Source Image Sequence | | All sequence and sub-sequence element cleared according to VRs | Primary |
| 0010,0050 | SQ | Patient's Insurance Plan Code Sequence | | All sequence and sub-sequence element cleared according to VRs | Primary |
| 0010,0101 | SQ | Patient's Primary Language Code Sequence | | All sequence and sub-sequence element cleared according to VRs | Primary |
| 0010,0102 | SQ | Patient's Primary Language Modifier Code Sequence | | All sequence and sub-sequence element cleared according to VRs | Primary |
| 0010,1002 | SQ | Other Patient IDs Sequence | | All sequence and sub-sequence element cleared according to VRs | Primary |
| 0038,0004 | SQ | Referenced Patient Alias Sequence | | All sequence and sub-sequence element cleared according to VRs | Primary |
| 0040,000B | SQ | Scheduled Performing Physician Identification Sequence | | All sequence and sub-sequence element cleared according to VRs | Primary |
| 0040,0275 | SQ | Request Attributes Sequence | | All sequence and sub-sequence element cleared according to VRs | Primary |
| 0040,0555 | SQ | Acquisition Context Sequence | | All sequence and sub-sequence element cleared according to VRs | Primary |
| 0040,1011 | SQ | Intended Recipients of Results Identification Sequence | | All sequence and sub-sequence element cleared according to VRs | Primary |
| 0040,1101 | SQ | Person Identification Code Sequence | | All sequence and sub-sequence element cleared according to VRs | Primary |
| 0040,4025 | SQ | Scheduled Station Name Code Sequence | | All sequence and sub-sequence element cleared according to VRs | Primary |
| 0040,4027 | SQ | Scheduled Station Geographic Location Code Sequence | | All sequence and sub-sequence element cleared according to VRs | Primary |
| 0040,4028 | SQ | Performed Station Name Code Sequence | | All sequence and sub-sequence element cleared according to VRs | Primary |
| 0040,4030 | SQ | Performed Station Geographic Location Code Sequence | | All sequence and sub-sequence element cleared according to VRs | Primary |
| 0040,4034 | SQ | Scheduled Human Performers Sequence | | All sequence and sub-sequence element cleared according to VRs | Primary |
| 0040,4035 | SQ | Actual Human Performers Sequence | | All sequence and sub-sequence element cleared according to VRs | Primary |
| 0040,A073 | SQ | Verifying Observer Sequence | | All sequence and sub-sequence element cleared according to VRs | Primary |
| 0040,A078 | SQ | Author Observer Sequence | | All sequence and sub-sequence element cleared according to VRs | Primary |
| 0040,A07A | SQ | Participant Sequence | | All sequence and sub-sequence element cleared according to VRs | Primary |
| 0040,A07C | SQ | Custodial Organization Sequence | | All sequence and sub-sequence element cleared according to VRs | Primary |
| 0040,A088 | SQ | Verifying Observer Identification Code Sequence | | All sequence and sub-sequence element cleared according to VRs | Primary |
| 0040,A730 | SQ | Content Sequence | | All sequence and sub-sequence element cleared according to VRs | Primary |
| 0070,0001 | SQ | Graphic Annotation Sequence | | All sequence and sub-sequence element cleared according to VRs | Primary |
| 0070,0086 | SQ | Content Creator's Identification Code Sequence | | All sequence and sub-sequence element cleared according to VRs | Primary |
| 0088,0200 | SQ | Icon Image Sequence | | All sequence and sub-sequence element cleared according to VRs | Primary |
| 0400,0402 | SQ | Referenced Digital Signature Sequence | | All sequence and sub-sequence element cleared according to VRs | Primary |
| 0400,0403 | SQ | Referenced SOP Instance MAC Sequence | | All sequence and sub-sequence element cleared according to VRs | Primary |
| 0400,0550 | SQ | Modified Attributes Sequence | | All sequence and sub-sequence element cleared according to VRs | Primary |
| 0400,0561 | SQ | Original Attributes Sequence | | All sequence and sub-sequence element cleared according to VRs | Primary |
| 4008,0111 | SQ | Interpretation Approver Sequence | | All sequence and sub-sequence element cleared according to VRs | Primary |
| 4008,0118 | SQ | Results Distribution List Sequence | | All sequence and sub-sequence element cleared according to VRs | Primary |
| FFFA,FFFA | SQ | Digital Signatures Sequence | | All sequence and sub-sequence element cleared according to VRs | Primary |
| 0008,0081 | ST | Institution Address | Data anonymised | | Primary |
| 0008,0092 | ST | Referring Physician's Address | Data anonymised | | Primary |
| 0008,2111 | ST | Derivation Description | Data anonymised | | Primary |
| 0018,A003 | ST | Contribution Description | Data anonymised | | Primary |
| 0040,0280 | ST | Comments on the Performed Procedure Step | Data anonymised | | Primary |
| 0040,1102 | ST | Person's Address | Data anonymised | | Primary |
| 0088,0906 | ST | Topic Subject | Data anonymised | | Primary |
| 4008,010B | ST | Interpretation Text | Data anonymised | | Primary |
| 4008,0300 | ST | Impressions | Data anonymised | | Primary |
| 4008,4000 | ST | Results Comments | Data anonymised | | Primary |
| 0008,0013 | TM | Instance Creation Time | 000000 | | Primary |
| 0008,0030 | TM | Study Time | 000000 | | Primary |
| 0008,0031 | TM | Series Time | 000000 | | Primary |
| 0008,0032 | TM | Acquisition Time | 000000 | | Primary |
| 0008,0033 | TM | Content Time | 000000 | | Primary |
| 0008,0034 | TM | Overlay Time | 000000 | | Primary |
| 0008,0035 | TM | Curve Time | 000000 | | Primary |
| 0010,0032 | TM | Patient's Birth Time | 000000 | | Primary |
| 0018,700E | TM | Time of Last Detector Calibration | 000000 | | Primary |
| 0038,0021 | TM | Admitting Time | 000000 | | Primary |
| 0040,0003 | TM | Scheduled Procedure Step Start Time | 000000 | | Primary |
| 0040,0005 | TM | Scheduled Procedure Step End Time | 000000 | | Primary |
| 0040,0245 | TM | Performed Procedure Step Start Time | 000000 | | Primary |
| 0000,1001 | UI | Requested SOP Instance UID | | UIDs are regenerated and remapped (original stored in local sqlite db) | Primary & Secondary |
| 0002,0003 | UI | Media Storage SOP Instance UID | | UIDs are regenerated and remapped (original stored in local sqlite db) | Primary & Secondary |
| 0004,1511 | UI | Referenced SOP Instance UID in File | | UIDs are regenerated and remapped (original stored in local sqlite db) | Primary & Secondary |
| 0008,0014 | UI | Instance Creator UID | | UIDs are regenerated and remapped (original stored in local sqlite db) | Primary & Secondary |
| 0008,0018 | UI | SOP Instance UID | | UIDs are regenerated and remapped (original stored in local sqlite db) | Primary & Secondary |
| 0008,0058 | UI | Failed SOP Instance UID List | | UIDs are regenerated and remapped (original stored in local sqlite db) | Primary & Secondary |
| 0008,010D | UI | Context Group Extension Creator UID | | UIDs are regenerated and remapped (original stored in local sqlite db) | Primary & Secondary |
| 0008,1155 | UI | Referenced SOP Instance UID | | UIDs are regenerated and remapped (original stored in local sqlite db) | Primary & Secondary |
| 0008,1195 | UI | Transaction UID | | UIDs are regenerated and remapped (original stored in local sqlite db) | Primary & Secondary |
| 0008,3010 | UI | Irradiation Event UID | | UIDs are regenerated and remapped (original stored in local sqlite db) | Primary & Secondary |
| 0008,9123 | UI | Creator-Version UID | | UIDs are regenerated and remapped (original stored in local sqlite db) | Primary & Secondary |
| 0018,1002 | UI | Device UID | | UIDs are regenerated and remapped (original stored in local sqlite db) | Primary & Secondary |
| 0020,000D | UI | Study Instance UID | | UIDs are regenerated and remapped (original stored in local sqlite db) | Primary & Secondary |
| 0020,000E | UI | Series Instance UID | | UIDs are regenerated and remapped (original stored in local sqlite db) | Primary & Secondary |
| 0020,0052 | UI | Frame of Reference UID | | UIDs are regenerated and remapped (original stored in local sqlite db) | Primary & Secondary |
| 0020,0200 | UI | Synchronization Frame of Reference UID | | UIDs are regenerated and remapped (original stored in local sqlite db) | Primary & Secondary |
| 0020,9161 | UI | Concatenation UID | | UIDs are regenerated and remapped (original stored in local sqlite db) | Primary & Secondary |
| 0020,9164 | UI | Dimension Organization UID | | UIDs are regenerated and remapped (original stored in local sqlite db) | Primary & Secondary |
| 0028,1199 | UI | Palette Color Lookup Table UID | | UIDs are regenerated and remapped (original stored in local sqlite db) | Primary & Secondary |
| 0028,1214 | UI | Large Palette Color Lookup Table UID | | UIDs are regenerated and remapped (original stored in local sqlite db) | Primary & Secondary |
| 0040,4023 | UI | Referenced General Purpose Scheduled Procedure Step Transaction UID | | UIDs are regenerated and remapped (original stored in local sqlite db) | Primary & Secondary |
| 0040,A124 | UI | UID | | UIDs are regenerated and remapped (original stored in local sqlite db) | Primary & Secondary |
| 0040,DB0C | UI | Template Extension Organization UID | | UIDs are regenerated and remapped (original stored in local sqlite db) | Primary & Secondary |
| 0040,DB0D | UI | Template Extension Creator UID | | UIDs are regenerated and remapped (original stored in local sqlite db) | Primary & Secondary |
| 0070,031A | UI | Fiducial UID | | UIDs are regenerated and remapped (original stored in local sqlite db) | Primary & Secondary |
| 0088,0140 | UI | Storage Media File-set UID | | UIDs are regenerated and remapped (original stored in local sqlite db) | Primary & Secondary |
| 0400,0100 | UI | Digital Signature UID | | UIDs are regenerated and remapped (original stored in local sqlite db) | Primary & Secondary |
| 3006,0024 | UI | Referenced Frame of Reference UID | | UIDs are regenerated and remapped (original stored in local sqlite db) | Primary & Secondary |
| 3006,00C2 | UI | Related Frame of Reference UID | | UIDs are regenerated and remapped (original stored in local sqlite db) | Primary & Secondary |
| 300A,0013 | UI | Dose Reference UID | | UIDs are regenerated and remapped (original stored in local sqlite db) | Primary & Secondary |
| 0010,21C0 | US | Pregnancy Status | 0 | | |